\pdfoutput=1
\documentclass[journal]{IEEEtran}

\usepackage{multirow}
\usepackage{graphicx}
\usepackage{amssymb,wasysym,marvosym}
\usepackage{threeparttable}
\usepackage[noadjust]{cite}
\usepackage[printonlyused]{acronym}
\usepackage[labelsep=newline,justification=centering,font={footnotesize,sc}]{caption}
\usepackage[hidelinks]{hyperref}
\usepackage{xcolor}

\graphicspath{{./img}}

\DeclareMathSymbol{*}{\mathbin}{symbols}{"03}

\begin{document}

\title{Energy-Efficient Softwarized Networks: A Survey}

\author{Iwan~Setiawan,~\IEEEmembership{Student~Member,~IEEE,}
        Binayak~Kar,~\IEEEmembership{Member,~IEEE,}
        and~Shan-Hsiang~Shen,~\IEEEmembership{Member,~IEEE}%
\thanks{Iwan Setiawan is with the Department of Computer Science and Information Engineering, National Taiwan University of Science and Technology, Taipei, Taiwan, and also with the Department of Electrical Engineering, Universitas Jenderal Soedirman, Purbalingga, Indonesia (e-mail: d10715810@mail.ntust.edu.tw; stwn@unsoed.ac.id).}%
\thanks{Binayak Kar and Shan-Hsiang Shen are with the Department of Computer Science and Information Engineering, National Taiwan University of Science and Technology, Taipei, Taiwan (e-mail: bkar@mail.ntust.edu.tw; sshen@csie.ntust.edu.tw).}}%

\maketitle

\begin{abstract}
With the dynamic demands and stringent requirements of various applications, networks need to be high-performance, scalable, and adaptive to changes. Researchers and industries view network softwarization as the best enabler for the evolution of networking to tackle current and prospective challenges. Network softwarization must provide programmability and flexibility to network infrastructures and allow agile management, along with higher control for operators. While satisfying the demands and requirements of network services, energy cannot be overlooked, considering the effects on the sustainability of the environment and business. This paper discusses energy efficiency in modern and future networks with three network softwarization technologies: SDN, NFV, and NS, introduced in an energy-oriented context. With that framework in mind, we review the literature based on network scenarios, control/MANO layers, and energy-efficiency strategies. Following that, we compare the references regarding approach, evaluation method, criterion, and metric attributes to demonstrate the state-of-the-art. Last, we analyze the classified literature, summarize lessons learned, and present ten essential concerns to open discussions about future research opportunities on energy-efficient softwarized networks.
\end{abstract}

\begin{IEEEkeywords}
Energy efficiency, software-defined networking, network functions virtualization, network slicing
\end{IEEEkeywords}

\section*{List of Acronyms}
\begin{acronym}[CAPEX]
\acro{5G}{Fifth Generation}
\acro{ALR}{Adaptive Link Rate}
\acro{AI}{Artificial Intelligence}
\acro{API}{Application Programming Interface}
\acro{BBU}{Baseband Unit}
\acro{BS}{Base Station}
\acro{CAPEX}{Capital Expenditure}
\acro{C-RAN}{Cloud Radio Access Network}
\acro{CPS}{Cyber-physical System}
\acro{CPP}{Controller Placement Problem}
\acro{CNF}{Cloud-native Function}
\acro{D2D}{Device-to-device}
\acro{DC}{Data Center}
\acro{DL}{Deep Learning}
\acro{DRL}{Deep Reinforcement Learning}
\acro{E2E}{End-to-end}
\acro{EC}{Edge Computing}
\acro{EDC}{Edge Data Center}
\acro{EE}{Energy Efficiency}
\acro{EM}{Energy Management}
\acro{EON}{Elastic Optical Network}
\acro{ETSI}{European Telecommunications Standards Institute}
\acro{FCAPS}{Fault, Config., Accounting, Performance, Security}
\acro{ICT}{Information and Communications Technology}
\acro{IoT}{Internet of Things}
\acro{KPI}{Key Performance Indicators}
\acro{MANO}{Management and Orchestration}
\acro{MEC}{Mobile/Multi-access Edge Computing}
\acro{ML}{Machine Learning}
\acro{MPLS}{Multiple Protocol Label Switching}
\acro{NFV}{Network Functions Virtualization}
\acro{NFVO}{NFV Orchestrator}
\acro{NOS}{Network Operating System}
\acro{NF}{Network Function}
\acro{NS}{Network Slicing}
\acro{NV}{Network Virtualization}
\acro{OEM}{Original Equipment Manufacturer}
\acro{OPEX}{Operational Expenditure}
\acro{OTN}{Optical Transport Network}
\acro{P-DP}{Programmable Data Plane}
\acro{QoS}{Quality of Service}
\acro{QoE}{Quality of Experience}
\acro{RA}{Resource Allocation}
\acro{RAN}{Radio Access Network}
\acro{RL}{Reinforcement Learning}
\acro{SDDC}{Software-Defined Data Center}
\acro{SDGs}{Sustainable Development Goals}
\acro{SDN}{Software-Defined Networking}
\acro{SDR}{Software-Defined Radio}
\acro{SFC}{Service Function Chaining}
\acro{TE}{Traffic Engineering}
\acro{UAV}{Unmanned Aerial Vehicle}
\acro{UWSN}{Underwater WSN}
\acro{VDC}{Virtual Data Center}
\acro{VM}{Virtual Machine}
\acro{VNE}{Virtual Network Embedding}
\acro{VNF}{Virtual Network Function}
\acro{WBAN}{Wireless Body Area Network}
\acro{WDM}{Wavelength-division Multiplexing}
\acro{WLAN}{Wireless Local Area Network}
\acro{WNV}{Wireless Network Virtualization}
\acro{WSN}{Wireless Sensor Network}
\end{acronym}

\section{Introduction}
\IEEEPARstart{S}{ustainability} remains an important issue, with the United Nations General Assembly aiming to complete Sustainable Development Goals (\acs{SDGs}) by 2030 \cite{united_nations_sustainable_2021}. As one of the topics, energy is prioritized to span sectors including \acs{ICT}. This sector contributed to more than 2\% of the total global production of the greenhouse gases, primarily CO\textsubscript{2}, in 2020 \cite{gesi_smarter2030, Belkhir_2018, ee_freitag_2021}. The good news is that ICT could reduce at least 20\% of global carbon emissions by 2030 \cite{gesi_smarter2030, sbt_guidance_2020}. However, this goal cannot be realized without effort by the global energy sector, whose target is to reduce CO\textsubscript{2} emissions to net-zero by 2050 \cite{noauthor_net_nodate}.

There is growing need to use connectivity and technology as fundamental components for realizing SDGs \cite{itu_broadband_2021}. With the increasing utilization of ICT, especially in the post-COVID19 era, energy costs will increase. The International Energy Agency (IEA) stated that \acs{DC}s and data transmission networks account for at least 2\% of global electricity consumption \cite{IEA2021}. Specific to the telecommunication industry, service providers have reported a need to spend 10\% or more of these costs, particularly for their \acs{OPEX} \cite{bell_gwatt_2015,mckinsey_greener_2020,gsma_ee_2019}. This phenomenon is not only happening at core networks, but also at the edge and access networks, where there has been significant growth of network devices and connections, including users with large numbers of smartphones, and IoT devices \cite{eenetw_bolla_2011,mckinsey_greener_2020,maaloul_energy_2018, gsma_intelligence_mobile_2021}.

There are two main reasons for saving energy in a sustainability context \cite{eenetw_bolla_2011, maaloul_energy_2018, bianzino_survey_2012}. The first is related to the environmental; the second is economical: \acs{CAPEX}, and OPEX in particular. Telecom operators have been trying to reduce these costs by investigating the financial damage and environmental issues, specifically carbon footprints. They found that by reducing carbon emissions, they can generate profit and come out ahead of the expenses invested to reach the goal \cite{maaloul_energy_2018}.

Research communities and industries have proposed techniques to deal with the issues of necessary energy savings. In the ICT sector, specifically computing, virtualization and cloud computing have been considered as the enablers to reach resource efficiency including energy \cite{Mastelic_2015} and sustainability \cite{gill_taxonomy_2019}. On the other hand, networking---the counterpart to computing---is also in the process of changing with so-called “network softwarization” to benefit from the promised programmability and flexibility features applied to network infrastructure.

Network softwarization can be described as the notion of designing, implementing, deploying, managing, and maintaining networks with the properties of software programmability \cite{AlDulaimi2018, afolabi_network_2018}. Several important technologies have been developed in order to realize this concept, primarily \acs{SDN} for softwarized network control, \acs{NFV} for virtualized network function, and \acs{NS} for isolated network resources. As a key component in softwarized networks, \acs{MANO} manages the life cycle of network resources and orchestrates them in single or multiple segments/domains. It also integrates and coordinates vertical/hierarchical network control and management planes for assuring isolated services \cite{afolabi_network_2018}. With this "glue" layer, operators would provide \acs{E2E} network services, guarantee the service levels, and achieve their objectives, e.g., QoS, security, or energy efficiency.

\subsection{Softwarized Network Scenarios}
We consider a softwarized network scenario as an environment---or a network setting---where a type of network is located in a segment or domain, with a particular network topology, traffic (flows), resources (compute, network, storage), and functions \cite{sos_he_2019} supported by softwarization technologies. Topology can be described as the arrangement of network components---i.e., basic network nodes such as switches and routers---and the interconnected links. Traffic or flows are related to the applications being run in the network, and resources are available in physical or virtualized forms with capacities and technologies specific to the scenario, e.g., optical fibers. Last, functions are defined as specialized network devices (physical or virtual) designed for particular purposes or applications. These include gateways, load balancers, intrusion detection systems, firewall, or related to a scenario, such as 5G \acs{VNF}s.

\begin{figure*}[h]
	\centering
	\includegraphics[scale=0.73]{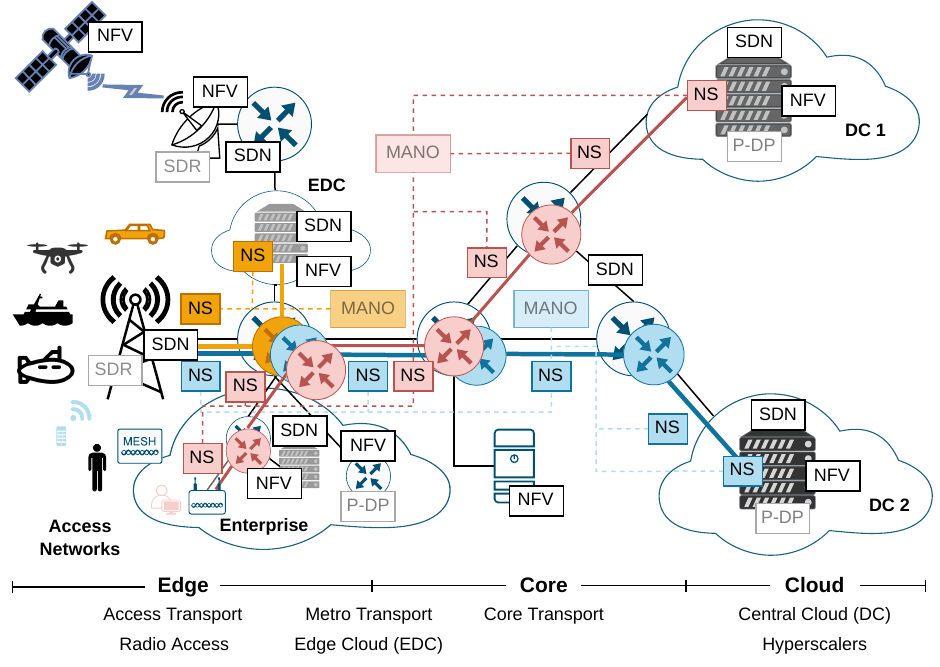}
	\caption{Network softwarization with various network scenarios located in different network segments.}
	\label{fig:netsoft}
\end{figure*}

Scenarios can be specific to a network segment or extended to a multi-domain network, each of which has characteristics. These scenarios can be grouped into categories, such as DC, transport, wireless, and emerging networks located in three segments: cloud, core, and edge. \figurename~{\ref{fig:netsoft}} illustrates network softwarization in various network scenarios from DC, transport to emerging networks at the edge, such as \acs{WBAN}, \acs{UWSN}, \acs{UAV}, vehicular, maritime, and satellite networks. As shown in the figure, different types of networks and distinct softwarization technologies can be implemented. The characteristics of each network scenario, as previously mentioned, has requirements that need to be satisfied, and will frame the softwarization solutions as well as the optimization approaches that can be used. In terms of energy efficiency, these unique requirements demand different energy consumption \cite{maaloul_energy_2018} with particular contributors, models, and strategies, which are described in Section \ref{section:ee-netw-infra}.

\subsection{Various Softwarized Network Scenarios}
Related to network segment, SDN in DCs (Cloud) is used for supporting computing tasks in controlling and monitoring DC resources. In this domain, SDN provides a global network view of the \acs{SDDC} and abstracts compute, storage, and network resources to be programmed by SDN controllers---which generally integrated also with DC controllers---based on actual and dynamic workloads. In Core segment, SDN has a role to program multiple high-speed network devices---commonly with optical technologies---to support interconnections among many nodes, which include networks in Cloud and Edge segments. In the latter domain, we also see a similar role of SDN, but with different edge and access technologies, e.g., passive optical or wireless networks. \acs{RAN} in wireless networks can be driven by an SDN controller, and \acs{SDR} for the radio parts. NFV utilized virtualization technology---and possibly cloud management--- to cater to the needs for flexible network functions in multiple domains. With VNFs which can be dynamically provisioned, network functions can be placed, chained, and consolidated locally, or remotely with migration techniques based on operators' objectives.

NS, as a network softwarization solution to provide E2E services, can be used as an example that utilizes multiple network softwarization technologies: SDN, NFV, also programmable data plane (\acs{P-DP}) and SDR. \figurename~{\ref{fig:netsoft}} shows a blue-colored network slice, i.e., a logical network with the required resources, spanning from a mobile phone on one end, to a base station, routers, and links to Cloud DC 2 on the other. Similarly, an orange-colored network slice, connecting from a vehicle to Edge DC (\acs{EDC}) shared the underlying physical network infrastructure with the blue-colored network slice. This is also applied to a pink-colored slice allocating network resources from a user in the Enterprise network to Cloud DC 1 through multiple network resources in transport networks located at the Edge and Core segments. These virtual networks will be dynamic according to the actual requirements for network services, orchestrated and optimized by MANO.

\subsection{Toward Energy-Efficient Softwarized Networks}
Compared with traditional networks, softwarized networks promise to provide flexible and scalable network management \cite{sos_he_2019}  and service provisioning. With these characteristics, networks with softwarization technologies such as SDN, NFV, NS, can be adaptive to different situations occuring in the network \cite{sos_kellerer_2019}. Dynamic adjustment or reconfiguration of resource allocation can be reached in a timely manner, either following traffic demands, security issues, or energy-efficiency objectives, e.g., reducing the number of live/operating network devices, or adapting node and link capacities according to the actual demands.

The adaptability, as we discussed previously, is difficult to apply in traditional networks, since these networks depend on the conventional control and management planes distributed in network nodes and functions. Energy efficiency techniques that are mainly applied in the physical or infrastructure layer, cannot be configured and adjusted directly/frequently based on actual situations or scenarios. To configure or reconfigure network devices or radio elements, networks typically use protocols such as the simple network management protocol (SNMP), with static configurations. Automating adjustment of devices to achieve an objective is also limited and tricky, such as introducing network monitoring applications with planned or defined scenario-based configurations. Moreover, implementation of the changes requires mechanisms of coordination of multiple network elements, which takes time, as there would be inconsistency of the network view. These approaches are considered difficult in the long term, and are not scalable with increasing numbers of devices, users, and running services; further, they are not flexible with different situations needed to be handled with fast and agile reconfiguration.

With the software properties adopted, a network infrastructure will be programmed and adaptively changed to dynamic situations, from serving different types of traffic with diverse applications to critical cases, such as defending from attacks. These include in stages of running a network infrastructure from design to operation, including reconfiguration to achieve an energy-saving objective. Thus, the questions are how softwarized network scenarios utilizing SDN/NFV/NS accommodate the goal in different network scenarios with energy-efficiency strategies, what kinds of attributes are considered in the literature, and what the challenges are arising from the state-of-the-art.

\subsection{Survey Methodology}
We divided our survey of energy-efficient softwarized networks into two parts. The first is survey-on-surveys to search available review journal papers covering (1) energy efficiency in networking; (2) network softwarization, including SDN, NFV, NS; and (3) energy-efficient softwarized networks. The second is the survey of each network softwarization technology for improving network energy efficiency.

We mainly used softwarized network scenarios as the base of the survey categorization, since energy efficiency differs from one scenario to the next. Related to network softwarization, each technology can be used differently in scenarios, and can be fitted to specific scenarios based on control/MANO layer (layer) and energy-efficient strategies (categories). Further, we categorize each reference to match with attributes of the solutions in classified papers by approach, evaluation, criteria, and metric.

\begin{figure}[h]
	\centering
	\includegraphics[scale=0.67]{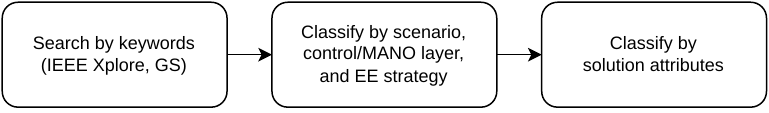}
	\caption{Survey steps for searching and classifying papers.}
	\label{search-classify}
\end{figure}

There are four considerations are included when categorizing research papers by network scenario: (1) an explicit scenario in the paper title; (2) a scenario implicit in the abstract and/or introduction; (3) an explicit scenario in the evaluation, based on the network topologies for the validation of algorithms or approaches; and (4) implicit or explicit scenarios in the discussion and/or in the conclusion sections.

\begin{figure}[]
	\centering
	\includegraphics[scale=0.38]{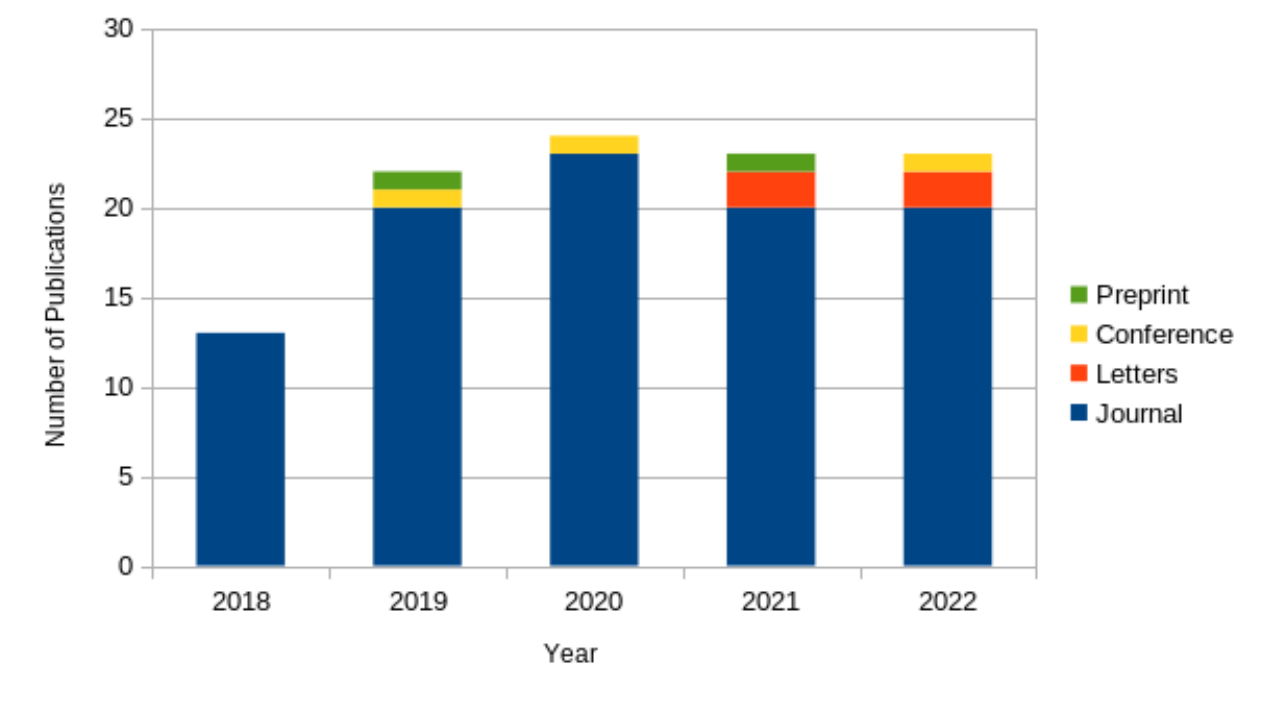}
	\caption{Number of publications per year used in the survey.}
	\label{fig:year-no}
\end{figure}

\subsubsection{Database Sources}IEEE Xplore was used as the primary database for searching the literature regarding energy-efficient softwarized networks. After several investigation attempts, we concluded that the topics of network softwarization and energy efficiency are mostly conducted by research communities in the venues covered by the database. However, we also utilized Google Scholar to find references from other databases such as ACM, Elsevier, Springer Nature, Hindawi, Wiley, IEICE, arXiv, and so on. With this approach, we can include the main venues of publications related to the topics and, at the same time, also incorporate relevant papers outside the primary source.

\subsubsection{Search Strategies}
\begin{figure}[h]
	\centering
	\includegraphics[scale=0.32]{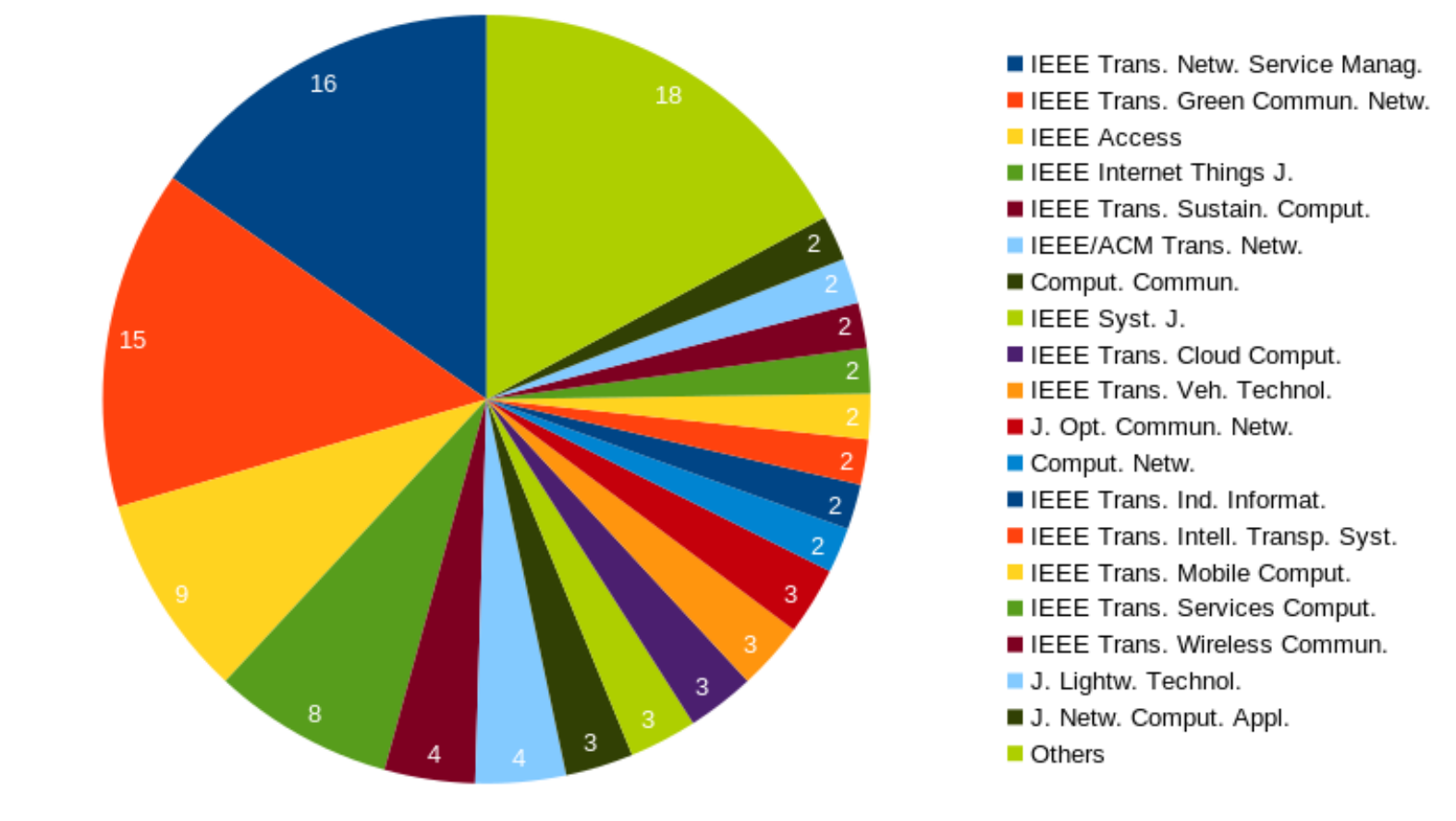}
	\caption{Articles per journal used in the survey.}
	\label{fig:chart-pub}
\end{figure}

After defining search strings for survey-on-surveys and the survey---along with each technology, problems, and related issues, we searched the literature from journal articles published in 2018 and onward (as per August 2022. Early-access articles were also included. Important survey papers prior to the year span were also added based on citation tracing. We selected references based on filters: title, abstract, and content focusing on upper layers of networking, i.e., logical. Since Google Scholar searching revealed huge numbers of results, we limited the search by 10 pages or included only 100 papers (but the relevant ones). These included relevant journal, conference, and magazine articles. We also marked reappeared papers, which were found in the main database, from the search results, assuming that these references would be more relevant to the search strings. \figurename~{\ref{fig:year-no}} shows the number of publications used in this survey. To complement this, \figurename~{\ref{fig:chart-pub}} illustrates the numbers of articles per journal.

\subsubsection{Classification}We categorized the literature found mainly by softwarized network scenario, control/MANO layer, and energy efficiency strategy. The next classification was determined by defined attributes of solutions, found in the literature, that will be described later in Section \ref{section:survey_ee_netsoft}. Lessons learned were summarized for each network scenario after analyzing the results of energy-efficient softwarized networks survey. Finally, challenges were classified by common future problems reported in multiple articles. Combining these with related sources, we formed the opportunities of future research directions on the topic.

\subsubsection{Survey Attributes}
\textbf{Approaches.} An approach is related to techniques that are used by researchers to obtain solutions. This comprises of exact, heuristic, scheme, policy, or no approach. We considered an approach to be classified to exact if it has a problem formulation or model run in a solver, e.g., Gurobi, CPLEX, GLPK. Heuristic approach was chosen if the solution has a heuristic to find a solution to a optimization problem in a certain time with approximation to the exact formulation.  A scheme is a system or combination of algorithms (can be heuristics), methods, procedures, protocols, information or other techniques in a specific environment to approach a problem. We consider a policy as a technique utilizing rules, guidelines, or priorities in the forms of intent to decide for solving a problem. \textbf{Criteria.} We categorized research on energy-efficient softwarized networks into five criteria. \textit{QoS} represents network performance that includes throughput, latency, jitter, and packet loss, as well as other user-oriented term like quality of experience (QoE). \textit{Scalability} is defined as a consideration to scale up or down network resources with methods and techniques applied to achieve it. \textit{Heterogeneity} is a criterion when a research deals with different types of elements, systems, or services. \textit{Mobility} is considered for moving components or resources in the network, either end-system, users, or logical elements, e.g., VNFs, virtual nodes or links. \textbf{Metrics.} The metric is comprised of energy, latency, and capacity. Some metrics, such as energy, have several submetrics that are grouped in energy, and include energy efficiency, energy/power consumption, residual energy, and lifetime. We only included references which have energy evaluation results, and not implicitly in costs. \textbf{Evaluation.} The evaluation method has two types: simulation and experimentation. For the later, the method is typically conducted in an emulation environment or testbed.

\subsection{Survey on Surveys}
In this subsection, we outline our survey-on-surveys regarding network softwarization and energy efficiency in softwarized networks, each of which consists of relevant review papers.

\begin{table}[]
	\centering
	\caption{Survey on Surveys on Energy Efficiency in Softwarized Networks}
	\label{table:es-netsoft}
	\begin{tabular}{ |p{0.5cm}|c|c|c|p{4.3cm}|}
		\hline
		\textbf{Ref.} & \textbf{SDN} & \textbf{NFV} & \textbf{NS} & \textbf{Focus of Survey} \\
		\hline
		\cite{rawat_software_2017} & \CIRCLE & \Circle & \Circle & Energy efficiency and security \\ \hline
		\cite{etengu_ai-assisted_2020} & \CIRCLE & \Circle & \Circle & Energy-aware TE, hybrid SDN, AI \\ \hline
		\cite{temesgene_softwarization_2017} & \CIRCLE & \Circle & \Circle &  Energy optimization tools, harvesting \\ \hline
		\cite{sos_assefa_2019} & \CIRCLE & \Circle & \Circle & Software-based methods and optimization models \\ \hline
		\cite{sos_thembelihle_2017} & \Circle & \LEFTcircle & \Circle & Sustainability with mobile NFs in 5G \\ \hline
		\cite{marzouk_energy_2020} & \LEFTcircle & \LEFTcircle & \LEFTcircle & Energy-efficient RA in 5G RAN \\ \hline
		{This work} & \CIRCLE & \CIRCLE & \CIRCLE & Energy efficiency in various softwarized network scenarios \\ \hline
	\end{tabular}
	\begin{tablenotes}[small]
		\item \CIRCLE Covered \LEFTcircle Partially covered \Circle Not covered
	\end{tablenotes}
\end{table}

Kellerer et al. \cite{sos_kellerer_2019} discussed adaptable softwarized networks with their potentials and challenges, including a framework that observes, composes, and controls functional primitives. Analysis by He et al. \cite{sos_he_2019} detailed flexibility as a common objective in softwarized networks and surveyed aspects, technologies, domains, and planes, in order to obtain a general understanding. \acs{5G} networks, as one of the softwarized network scenarios, were reviewed in \cite{lake_softwarization_2021}, where enablers, platforms, and tools for realizing softwarization in 5G were considered, along with standardization activities. In a similar scenario, Khan et al. \cite{khan_survey_2020} discussed security and privacy issues in relation to the technologies. Softwarization in mobile edge cloud targeting tactile internet use case is considered in \cite{sos_cabrera_2019}.

Table {\ref{table:es-netsoft}} shows a comparison of studies on energy efficiency in softwarized networks, along with their attributes related to SDN/NFV/NS. As seen in the table, some studies discussed energy efficiency with network softwarization. However, this work has predominantly focused on one technology (e.g., SDN), or has been limited to a specific network scenario. Rawat and Reddy \cite{rawat_software_2017} analyzed energy efficiency, security, and their trade-offs related to SDN. Survey on energy-aware \acs{TE}, hybrid SDN, and the potentially beneficial role of \acs{AI} has been conducted in \cite{etengu_ai-assisted_2020}. Temesgene et al. \cite{temesgene_softwarization_2017} examined the paradigm of softwarization, energy harvesting, and optimization techniques aiming to achieve system requirements in future mobile networks. Assefa and Ozkazap \cite{sos_assefa_2019} discussed software-based methods and optimization models for energy-efficient SDNs. The NFV in 5G mobile networks scenario is reviewed in \cite{sos_thembelihle_2017}, targeting on energy-efficient 5G architectures using mobile \acs{NF}s. At last, Marzouk et al. \cite{marzouk_energy_2020} investigated resource allocation for the same network scenario, focusing on energy efficient \acs{RA} in shared RANs within a multi-operator context.

\subsection{Contributions and Paper Organization}
This paper contributes to the body of knowledge regarding energy efficiency in three main areas of network softwarization: SDN, NFV, and NS. The contributions of this paper can be summarized as follows.

\begin{enumerate}
	\item The contributors of energy consumption in network infrastructure, energy consumption models, and the taxonomy of energy-efficiency strategies for obtaining green networking are described and proposed.
	\item The basic principles and architectures of SDN, NFV, and NS are reviewed and put into an energy efficiency context highlighting control and MANO layers as key components to reach the objective.
	\item The survey results, including classification and taxonomy based on softwarized network scenarios, control/MANO layer, energy-efficiency strategy, with approach, evaluation method, criterion, and metric attributes, plus a short summary of observation related to the strategy, SDN/NFV/NS problem(s), and subscenario/application/other details, e.g., a particular ML scheme.
	\item Finally, the important points captured from the survey were explored and analyzed, and possible future challenges are presented.
\end{enumerate}

\begin{figure}[h]
	\centering
	\includegraphics[scale=0.83]{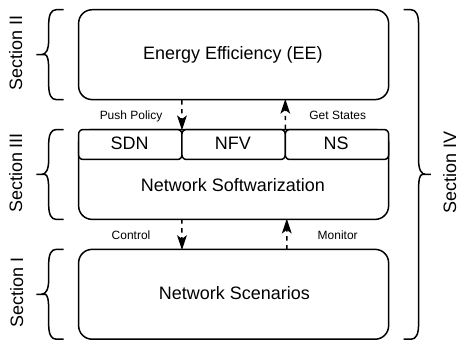}
	\caption{Main topics discussed in the paper.}
	\label{fig:outline}
\end{figure}

\figurename~{\ref{fig:outline} shows the main topics discussed in this paper. It} is divided into six sections. Section \ref{section:ee-netw-infra} describes energy efficiency in network infrastructure, including energy-consumption contributors, models, energy-efficiency strategies to achieve green networking, energy efficiency and network softwarization---in relation to virtualization and cloud computing. In Section \ref{section:netsoft}, we provide brief detail on SDN, NFV, and NS emphasizing control and MANO layers for obtaining network energy efficiency. Section \ref{section:survey_ee_netsoft} includes the survey on energy efficiency in softwarized networks on each technology, based on network scenarios, control/MANO layer, and categories of energy-efficiency strategies, with classification on approach, evaluation, criterion, and metric attributes of each solution, including a summary of our observation regarding the strategy, specific softwarization problem, and subscenario/application/other details is discussed, along with insights obtained and analyzed. We outline possible future research directions in Section \ref{section:frd}. Finally, concluding remarks for the literature review are delineated in Section \ref{section:conclusions}.

\section{Energy Efficiency in Network Infrastructure}
\label{section:ee-netw-infra}
Based on the environmental perspective, "green networking" is a term used to define efforts to reduce greenhouse gas emissions in network infrastructure. There are other perspectives as well, economical, regulatory, and engineering ones. The latter is defined as a mechanism to decrease the energy consumption in running a job while maintaining a similar degree of performance \cite{bianzino_survey_2012}. We use the engineering perspective as the main consideration for environmental and economical effects of the applied techniques or strategies.

\subsection{Energy Consumption Contributors}
\label{subsection:ec-contrib}
There are four general elements in network infrastructure---wired and wireless networks---that contribute to energy consumption: networked hosts, network devices, network architecture, and network services \cite{gesi_smarter2030, bianzino_survey_2012, buzzi_survey_2016, maaloul_energy_2018}.

\subsubsection{Networked  hosts}
Computers or other types of end systems are the consumers/clients or producers/servers of network services provided by network application, such as servers in cloud or edge DCs, and smartphone or IoT devices connected to access networks at the edge. The area where these end systems reside is considered well-developed \cite{bianzino_survey_2012}, as they are mainly derived from computing systems. Hosts will have both hardware and software, such as operating systems and applications, which consume energy and affect energy consumption. In wireless networks, end systems can be laptops, mobile phones, tablets, IoT devices, or other appliances.

\subsubsection{Network devices}
In the network infrastructure, network devices are the second most common contributors after networked hosts. In a network domain, these components affect the energy consumption profile for each scenario with a distinct numbers of devices \cite{maaloul_energy_2018}. Every device, such as network switches and routers, has unique characteristics, such as the number of active ports, enabled protocols, and traffic. These would also apply to network functions---specialized network devices for specific applications or services. For wireless networks, 5G in particular, base stations along with radio frequency transceivers are considered the most energy consumers \cite{Lopez_Perez_2022}. Improvements for “greening” network devices or nodes are available through hardware and software, similar to networked hosts.

\begin{figure}[]
	\centering
	\includegraphics[scale=0.64]{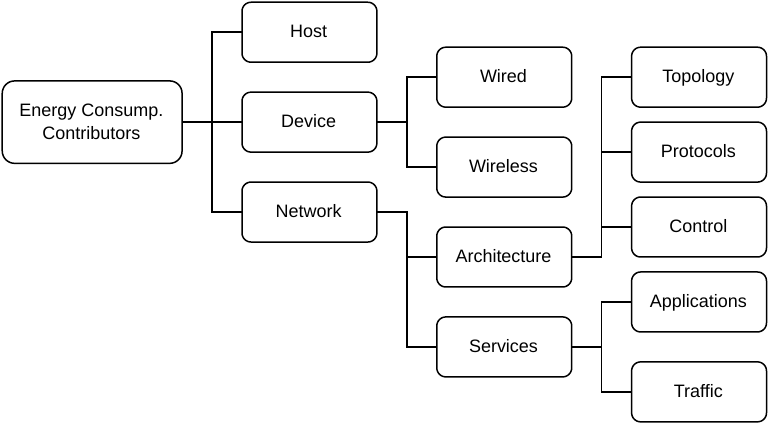}
	\caption{Energy consumption contributors in network infrastructure.}
	\label{fig:ec-contrib}
\end{figure}

\subsubsection{Network architecture}
The architecture is related to network design and consists of network devices and how they are connected and configured. Network architecture is generally divided into three domains: core, metro or transport, and access networks \cite{maaloul_energy_2018}; each is related to network topology and has scalability considerations. The use of optical fibers as the medium contributes to energy consumption in core and transport networks. However, access networks are typically in the form of wireless networks.

\subsubsection{Network services}
Energy consumption in network infrastructure is not only influenced by networked hosts, devices, and the architecture, but also by services run on the network. The protocols and traffic characteristics of the applications are considered here. Energy consumption in network infrastructure is also affected by these services and depends on the characteristics of the applications (e.g., routing, TE, security, and multimedia) and how they used the physical network resources to provide services to the users or end systems.

\subsection{Energy Consumption Models}
Energy consumption models are needed in different network scenarios since each of which has diverse networked hosts, network devices, and network characteristics. The models can be defined as energy or power models. These depends on the time unit, i.e., time period in seconds or unit times. If the time unit is in the latter, energy (Joules) and power (Joules/second) can be considered equivalent \cite{ec_dayarathna_2016}. Researchers either choose energy or power for different quantities but similar objective, i.e., energy efficiency or power efficiency. Since the specifications of network components, including hosts and devices, are in power quantity, we will use power to describe energy consumption models with the assumed time unit.

Common characteristics when modeling energy/power consumption in different scenarios include static and dynamic components. The static power consumption is the base power consumption, for example power that is consumed by servers or switches when they switched on. The dynamic can be defined as load-dependent parameters and the quantities change when workloads or traffic loads become fluctuate.

\subsubsection{Networked Hosts}
Networked hosts, for example servers in a DC, has two common characteristics: static and dynamic power consumption. Generally, modeling hosts uses their functional components, i.e., processor, memory, storage, network interface card, etc. Thus, to get a linear power consumption model, we need to sum all of the power consumptions from those components to get the total power consumption of a server. The model considered as additive server power models \cite{ec_dayarathna_2016}. The other models are utilizing frequency used by processor in a server. There are additive power consumption quantities from components except the processor, and the other one is frequency in which the processor is running and a constant representing capacitive dynamic power consumption. Details of these models including cloud servers can be read on the references \cite{ec_dayarathna_2016, ec_lin_2020}.

\subsubsection{Wired Network Devices}
In wired network devices, e.g., switch and routers, we have main subsystems to be considered for power modeling, namely control plane, data plane, and environmental \cite{ec_dayarathna_2016}. The data plane subsystem is considered as load-dependent. Thus, it has further power modeling with three components: power consumption when idle, processing packet, and store-and-forward. Specific to optical networks, there are major elements for power modeling in core networks. Since these networks are multilayer with IP, Ethernet, \acs{OTN}, and \acs{WDM}, the sum of the power consisting of the layers. These parameters are further detailed to account power consumption of optical switches, transponders, amplifiers, and regeneration \cite{ec_dayarathna_2016}.

\subsubsection{Wireless Network Devices}
Energy consumption in a base station primarily draws from power amplifier and signal processing circuits \cite{ec_niu_2021}. Power amplifier model is based on bandwidth, power amplifier efficiency, total transmit power, and static energy consumption. For signal processing circuit, the model comes from radio frequency chain considering the number of antennas. For end-user device (LTE), the power model includes sending and receiving data with a connected state \cite{ec_hoyhtya_2018}. Each of which has power consumption of cellular and baseband subsystems, transmission or receiving power, and additional power for being on or active.

\subsection{Energy-Efficiency Strategies}
Sustainability is a more general consideration than energy problems. Related to that, "green networking" is used by industries and researchers to refer to efforts to achieve energy efficiency and reduce CO\textsubscript{2} footprints of network infrastructures. Derived from the literature \cite{eenetw_bolla_2011, buzzi_survey_2016, Lopez_Perez_2022}, reducing network energy consumption for achieving green networking has six strategies: hardware-based improvements, dynamic adaptation, sleep modes, network heterogeneity, energy heterogeneity, and machine learning.

Table {\ref{table:ee-strategies}} shows the strategies and relation with the affected elements in networking, which are described as energy consumption contributors in subsection \ref{subsection:ec-contrib}. These categories will be used later in Section \ref{section:survey_ee_netsoft} as one of the main attributes.

\subsubsection{Hardware-based Improvements}
Hardware-based improvements consist of re-engineering hardware components and designs from semiconductor to device, including processor and memory as two main elements in host and network devices. This also apply to software that is embedded and run on the hardware. Related to wireless networks, base stations in particular, there are components that can be re-engineered to improve their energy efficiency. For instance, power amplifiers, since these elements will affect transmission power of a base station. Because this strategy is related to the hardware, we do not include this category for classification in Section \ref{section:survey_ee_netsoft}.

\subsubsection{Dynamic Adaptation (DA)}
This strategy includes two groups: hardware-based and network-based. In hardware-based DA, three strategies are for the processor: dynamic voltage scaling (DVS), dynamic frequency scaling (DFS), or both (DVFS), and efficient used of memory or ternary content-addressable memory (TCAM). All of these related to power management feature in computing systems. Network-based DA uses adaptive link rates (\acs{ALR}) to switch between supported rates of network interfaces or bundled links on network devices. For wireless networks, there is similar adaptation related to radio transmission power to adjust coverage with cell zooming. The network-based DA can be used to consolidate traffic to certain network devices or path, so that low utilized network devices can be set to lower rate or switch to idle or sleep. For the later, this technique combines DA and SM, discussed in the next point. In short, DA is based on the computing load for the hardware-based and traffic load for the network-based. With this strategy, network devices, also networking, can be scaled up and down as long as they are supported by the hardware. Networking with network softwarization can be flexibly scaled up and down, and/or consolidated (e.g., by re-routing traffic to specific path to reduce active nodes, or less rates of certain devices). With virtualization, for example with NFV or NS, virtual functions or networks can be consolidated or migrated to achieve energy efficiency. Thus, it will change the topology in terms of control and control messages and, more importantly, network resources. The other techniques are related to storage are data compression and efficient used of TCAM. In general, DA is dealt with dynamic resource provisioning including control and allocation. The objective can be performance so that the network configuration will be set to support high performance, or energy efficiency. These two objectives commonly need to be considered for operators, since the changes in one will affect to the other. Thus, the trade-offs between the two demand new approaches, e.g. multi-objective optimization, ML schemes.

\begin{table}[]
	\centering
	\caption{Categories of EE Strategies}
	\label{table:ee-strategies}
	\begin{tabular}{ |p{3.8cm}|c|c|c|}
		\hline
		\textbf{Category} & \textbf{Host} & \textbf{Device*} & \textbf{Network} \\
		\hline
		Hardware-based Improvements & \CIRCLE & \CIRCLE & \Circle \\ \hline
		Dynamic Adaptation (DA) & \CIRCLE & \CIRCLE & \CIRCLE \\ \hline
		Sleep Modes (SM) & \CIRCLE & \CIRCLE & \CIRCLE \\ \hline
		Heterogeneous Network (HT) & \Circle & \CIRCLE & \CIRCLE \\ \hline
		Energy Harvesting (EH) & \Circle & \CIRCLE & \Circle \\ \hline
		Machine Learning (ML) & \Circle & \Circle & \CIRCLE \\ \hline
	\end{tabular}
	\begin{tablenotes}[small]
		\item * Including wired and wireless, e.g. switches, base stations.
	\end{tablenotes}
\end{table}

\subsubsection{Sleep Modes (SM)}
This strategy is inline with general energy saving techniques by switching off physical devices to reduce energy consumption. With multiple modes (e.g., active, sleep, idle) network devices can consume less energy. It cannot be realized without hardware support for compute, memory, and network components, i.e., hardware-based improvements from the manufacturers or involving external small devices that take over the connection to the network for the main device (proxy) \cite{eenetw_bolla_2011}. Components that can be set to sleep modes are device, line cards, and ports in wired network devices. In wireless networks, base stations can be switched off dynamically. The modes can be micro-sleep (up to 160 ms for bursty traffic) or macro-sleep (minutes to hours based on the distribution of traffic and demands) \cite{Lopez_Perez_2022}.

\subsubsection{Heterogeneous Network (HT)}
Heterogeneous network resources can be used to improve energy efficiency, since different scenarios (e.g., traffic or compute loads) can use specific resources efficiently. Moreover, if we consider distance or latency, the closer a network device, the more efficient it is for users or end systems to connect with. This scenario would be true, especially in wireless networks, where transmission power of base stations would consume more power. Heterogeneous networks (HetNet) provide a way to combine one or small numbers of macro-cell with multiple small devices, instead of using multiple macro-cells, which would consume higher energy if not configured correctly (this technique can be a source of interferences and, thus, consume more energy). However, heterogeneous network resources demand good network management, since there are many physical differences of devices and they have distinct characteristics to be managed for resource control and allocation.

\subsubsection{Energy Harvesting (EH)}
Energy harvesting has its meaning in wireless networks, mobile or cellular in particular. This strategy is used to obtain or harvest energy from the environment (ambient sources), e.g., radio frequencies or signals, and renewable energy such as wind and solar \cite{buzzi_survey_2016}. The latter, commonly called green energy, is included as the source for energy harvesting techniques. We extend the definition of energy harvesting to the general techniques for harvesting energy from ambient and renewable energy sources. Thus, it is not only specific to the term in green wireless networking. The characteristic of using these sources of energy is the amount of energy that can be harvested. However, with data for wind and solar, also calculation of wireless network planning with dense networks and MIMO, this strategy can be realized with prediction via traditional data mining or machine learning techniques.

\subsubsection{Machine Learning (ML)}
Energy efficient strategies, such as DA and SM, will improve in topology-oriented and traffic-oriented scenarios. With dynamic situations, various service requirements and heterogenous network resources, there will be limitations for applying the techniques, particularly in real environment. Scalability and complexity are the two main problems in real scenarios with parameters, stochastic and non-linear characteristics of the processess \cite{Lopez_Perez_2022}. Therefore, machine learning comes as one of the strategies to tackle these problems by providing intelligence especially for network operations. By using network data, obtained from measurements, supervised and unsupervised learning will model the behaviour of the network and decide/predict the next actions. With less network data collected and unknown situations, RL is adopted by using agents interacting with the environment to obtain high rewards, for example finding near-optimal DA or SM policies to make the network energy consumption minimum.

\subsection{Energy Efficiency and Network Softwarization}
Virtualization is the "softwarization" in computing by abstracting the physical hardware, so that \acs{VM}s or containers can be run on top of the virtualization layer. With the abstraction layer, virtualization consolidates multiple environments with different operating systems and applications run on top of hardware or physical resources. This process can lead to efficiency of resources, including energy. Cloud computing, on the other hand, scales virtualization to higher levels with stricter requirements (such as service-based, on-demand, elastic, measured services \cite{Mastelic_2015}) in pools of high volume but generic servers stacked in racks within DCs owned by enterprises or hyperscalers. As DCs are consuming high amounts of energy, virtualization and "cloudification" give enterprises and providers a way to manage that consumption, by consolidating and orchestrating huge resources in a DC or multi-DC, with either private or public cloud DCs (multi-cloud).

Complementary to computing, networking is the next part of the evolution in order to adopt software programmability properties into the network. The benefits of network softwarization, by abstracting or virtualizing network infrastructure into logical networks with required resources, would provide flexibility in provisioning, reconfiguring, and deactivate resources, also scaling and consolidating them. This characteristic will make the network adaptable to dynamic situations, manageable to heterogeneous resources and varying services. Moreover, with different requirements and various objectives of the operators, the advantages are obvious, in terms of costs or sustainability for businesses and environment. In addition to that, the promises of network softwarization for giving more control and management to the abstraction or virtualization of the network infrastructure, will drive further. Not only in managing diverse network resources in different segments, but also in orchestrating multiple control and management layers. Thus, MANO layer will be an utmost important component to integrate and optimize all of the physical and virtual resources in support of providing sustainable services.

\section{Network Softwarization: The Key Enablers}
\label{section:netsoft}
Three technologies of network softwarization, i.e., SDN, NFV, and NS, will be described in this section, along with MANO as a key element to manage and orchestrate life cycle of a network infrastructure (physical and virtual resources) for providing E2E services. Each technology will be characterized with principles, architecture, and example use cases of energy-efficient strategies applied in softwarized networks.

\subsection{Software-Defined Networking}
Software-defined networking (SDN) is a network architecture that has four main principles \cite{sdn_kreutz_2015}, which grew out from the basic ideas of computing, abstraction in particular. These foundations are:
\begin{enumerate}
	\item abstraction of networking planes so that the control plane is physically separated from the \textit{data plane} \cite{onf_sdn};
	\item SDN controller as network operating system running at the \textit{control plane}, which manages the underlying resources of the data plane;
	\item \textit{management plane} with software modules or applications driving and controlling the network through one SDN controller or more; and
	\item match-and-action forwarding decisions based on packet flow, implemented via APIs \cite{onf_sdn}, e.g., OpenFlow.
\end{enumerate}

\begin{figure}[]
	\centering
	\includegraphics[scale=0.72]{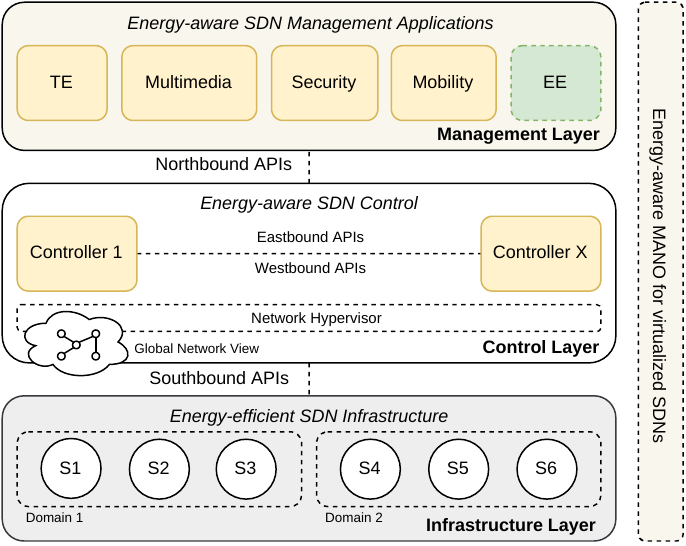}
	\caption{General SDN architecture with energy efficiency context.}
	\label{fig:sdn}
\end{figure}

\figurename~{\ref{fig:sdn}} shows SDN architecture in general with the additional context of network energy efficiency. We use infrastructure, control, and management layers to refer to data, control, and management planes. In this figure, control and management planes are important layers for achieving energy-efficient SDN infrastructure (data plane), along with MANO for orchestrating the resources, and integrating control and management layers. Since one or more SDN controllers are logically centralized, they store global network view of the network infrastructure so that the network topology, flows, and states of resources are known. This feature will support efficient use of network resources and the mechanisms to exchange the information among controllers and controllers-to-devices by way of eastbound/westbound and southbound APIs respectively.

For the case of multi-domain networks---where there are multiple network segments---operated by one or more providers, the control and management layers can be orchestrated to provide E2E services with coordinated control planes and integrated management by use of MANO. In this matter, optimization process, e.g., with an energy efficiency objective, can be run in management or MANO layer. The operators can use DA, SM, or other strategies. For routing or TE application, the traffic can be re-routed to the most energy-efficient path, and unused network devices can be set to idle or switched off. In the case of low traffic with best-effort requirement, the speed of the network interface or line card can be set lower with the help of ALR.

\subsection{Network Functions Virtualization}
\acs{ETSI} has been developing standards regarding NFV. This technology is defined as a method to create dynamic and service-oriented networks targeting development cycles of products, as well as CAPEX and OPEX, while providing improved services \cite{mijumbi_network_2016}. NFV increases resource efficiency by replacing network functions/appliances with software environments in the forms of VMs or containers on top of generic or commodity hardware.

In a recent development, container-based virtualization is preferred, since it is lightweight, has lesser use of memory and storage resources, minimum overheads, and can be quickly migrated \cite{nfv_attaoui_2023} and orchestrated with cloud-based management tools such as Kubernetes. Based on operating system-level virtualization (rather than hardware virtualization), containers are in small packages that include applications providing microservices (usually in the form of clusters). With these characteristics, containerized or cloud-native functions (CNFs) are considered better in terms of energy consumption compared to VNFs, since they provide efficient use of computing and networking resources while also retaining performance.

\begin{figure}[]
	\centering
	\includegraphics[scale=0.72]{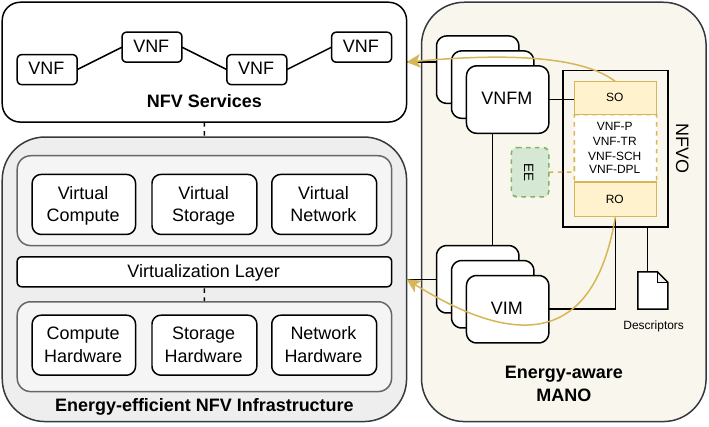}
	\caption{NFV architecture with energy-aware MANO.}
	\label{fig:nfv}
\end{figure}

The NFV architecture based on ETSI \cite{etsi_nfv_2013} has three important components, namely, NFV infrastructure (NFVI), NFV Services with \acs{VNF}s, and MANO as described in \figurename~{\ref{fig:nfv}}. NFVI provides the needed infrastructure for NFV with computing, storage, and network hardware at the bottom layer. On top of that, the virtualization layer with hypervisor will virtualize the physical infrastructure to create environments with virtual compute, storage, and network resources. The role of NFV MANO (\acs{NFVO}) is to orchestrate NFVI resources and manage VNFs life cycle with the help of VIM and VNFM. As shown in the figure, MANO is the component driving the NFVI, VNFs, and \acs{SFC} to be energy-oriented, similar to the function of SDN control plane.

We consider NFVO into two parts: \textbf{resource orchestration (RO)} and \textbf{service orchestration (SO)}. We derived these two from the ETSI OSM \cite{nfv_yi_2018}. RO is dealt with physical/virtual resources provisioning to particular services, and SO handles with instantiation of VNFs to provide and orchestrate NFV services. With these two components in NFVO, operators can run algorithms related to VNF placement (VNF-P), traffic routing/steering (VNF-TR), scheduling (VNF-SCH), deployment (VNF-DPL) \cite{nfv_yi_2018}, and reconfiguration (VNF-RCF) based on the optimization targeting energy efficiency.

For our classification in the next section, RO covers VNF-P, VNF-SCH, VNF-DPL, and VNF-RCF, while SO comprises VNF-TR. The latter includes chain composition (CC) and function graph embedding (FGE) \cite{nfv_gil-herrera_2016}. Related to NFV services, SFC is defined as VNF chaining or routing that consists of multiple VNFs to provide services. Since SFC focuses on network services and it includes VNF-P and VNF-TR, we classified SFC to network slicing in network service virtualization (SV).

\subsection{Network Slicing}
Network virtualization (NV) is not a new technology in networking, as forms of this, such as virtual local area networks (VLAN) and more advanced technology, i.e., \acs{MPLS}, have been in existence for some time. The earlier is used for creating network slices at the link layer, while the later is for forwarding tables \cite{sos_blenk_2016}. With the advent of 5G mobile networks, the “new” and popularly discussed term is "network slicing" (NS). The concept of NV is mainly overlay networks in traditional networks, defined as how a physical network can be shared or divided into virtual networks on top of the network infrastructure. Platforms to provide NV started to emerge with the well-known PlanetLab \cite{ns_chun_2003, Barakabitze2020}. Extending the NV concept, NS arised in the context of 5G networks with E2E systems \cite{boutaba2021managing}, with radio access, transport, and core networks.

\begin{figure}[]
	\centering
	\includegraphics[scale=0.54]{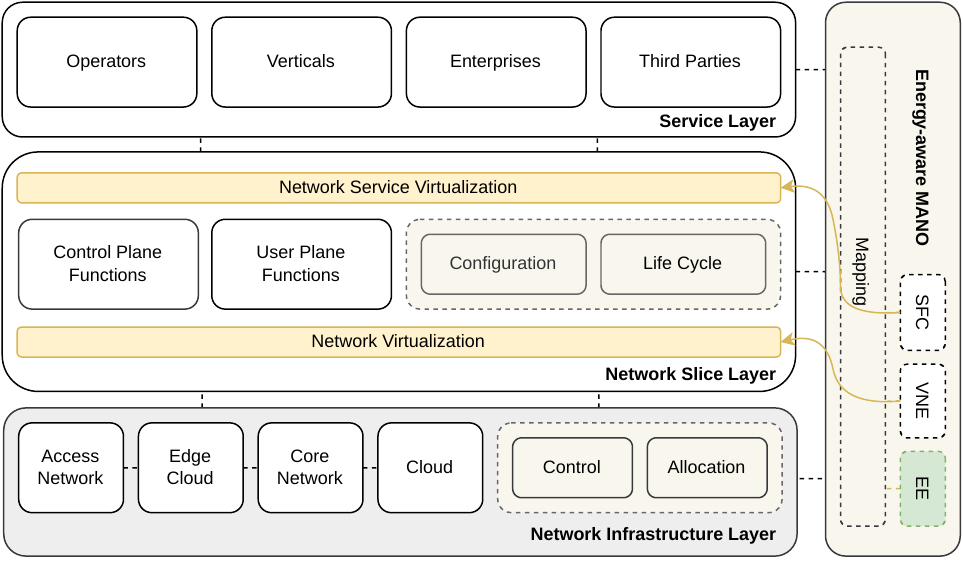}
	\caption{Network slicing architecture with energy-aware MANO.}
	\label{fig:ns}
\end{figure}

The architecture of NS, based on Next Generation Mobile Networks Alliance (NGMN), consists of resources, network slice instances, and service slice instances, with infrastructure, virtual networks, and services at the top, but without the MANO component. \figurename~{\ref{fig:ns}} describes an NS architecture with energy-aware MANO, adapted from \cite{Foukas_2017}. Identical to the SDN controller and NFV MANO, NS MANO has important functions for managing and orchestrating network resources. With energy-aware MANO, the network virtualization and network function layers will be provisioned over the infrastructure layer, fulfilling the requests of operators, verticals, enterprises, or third parties.

We consider two big problems represented by two virtualization layers. First, \textbf{network virtualization (NV)} for dealing with virtualizing lower physical infrastructure to provide virtual networks that consist of wired and wireless. This lower virtualization mostly based on VNE applied to lower level of the layers. Second, \textbf{service virtualization (SV)} for dealing with a layer that provides isolated slices for various services from third parties, operators, enterprises, etc. The problems in SV are related to dynamic reconfiguration of VNF and SFC (VNF-P and VNF-TR) in multiple slices, joint stages of NFV-RA, and joint resources of compute, network, storage or caching spanning from core to edge networks. These two problems commonly use graph- and set-based solutions.

\subsection{Management and Orchestration}
As observed in the previous subsections, the role of MANO is key in end-to-end service management with multi-domain, multi-operator, and multi-tenant using the combinations of network softwarization technologies. Networking resources, along with computing and storage/caching, need to be managed with each of the controller in every domain, but we need MANO to create an efficient and flexible infrastructure holistically by managing and orchestrating all the resources from multiple domains. With this kind of technique, MANO provides an abstraction to the deployment and operations of network services \cite{mano_desousa_2019}. Related to energy efficiency, MANO can coordinate resource management to obtain the objective and, in most cases, balance with other objectives such as QoS and security. There would be a need to do the computation and provide the results to the service orchestration functions managing multiple resources end-to-end.

There are some open source implementations of MANO, contributed by various organizations. De Sousa et al. \cite{mano_desousa_2019}, for example, classified the features ranging from VNF definition, resource domain, elements of MANO, interface management, to multiple domains. Based on the results, ONAP \cite{onap} and OSM \cite{osm} have potential to be used for MANO since they provide all the features, except OSM, which has one drawback for use in multiple domains.

\section{Survey on Energy-Efficient Softwarized Networks}
\label{section:survey_ee_netsoft}
We classified the literature on energy efficiency in/with network softwarization mainly based on network scenario, control/MANO layer, and categories of energy efficiency strategies. These include five major scenarios: DC, transport, wireless, and emerging networks. We added a new scenario if the numbers of articles found in a specific scenario (subscenario) was considerable, e.g., \acs{WSN} in energy-efficient SDN.

Along with the network scenario, categories of energy efficiency strategies were used to further categorize the literature. As previously described in Section II, we have five strategies to be considered: DA, SM, HT, EH, and ML. The control/MANO layer, i.e., SDN: control or management; NFV: RO or SO; NS: NV or SV. As an additional field,  short summary of observation related to specific green strategy, SDN/NFV/NS problem(s), and subscenario/application/other details, e.g., a particular ML scheme. After classifying ongoing research papers based on network scenario, control/MANO layer(s), and energy efficiency strategy, we continued the process with attributes related to each of them, as follows. We mark the respected research by using considered, partially considered, and not considered. Regarding the partially considered, we defined this to be related to physical resources, e.g., heterogeneity marked considered if the authors consider physical resources.

\subsection{Energy-Efficient SDN}
The survey on energy-efficient SDN featured a considerable amount of papers, including studies of emerging network scenarios, such as IoT, satellite, UAV, and vehicular networks, also WBAN, UWSN, and edge cloud (EDC). Only papers representing each subscenario were included. Since WSN scenario has multiple articles found and classified, we created separated scenario to gain the specifics. Based on network scenario, we detailed each of the articles with respect to its attributes that were previously defined and compared in the beginning of Section \ref{section:survey_ee_netsoft}. The results are shown in Table {\ref{table:es-sdn}}.

\begin{table*}
\centering
\caption{Survey on Energy-Efficient SDN \\ (\Circle Not considered \LEFTcircle Partially considered \CIRCLE Considered)}
\label{table:es-sdn}
\begin{tabular}{|c|p{0.55cm}|p{0.6cm}|p{0.5cm}|p{0.2cm}|p{0.2cm}|p{0.2cm}|p{0.2cm}|p{0.2cm}|p{0.2cm}|p{0.2cm}|p{1.1cm}|p{0.6cm}|p{6.5cm}|}
\hline
\multirow{7}{*}{\rotatebox{90}{\textbf{Netw. Scenario}}} & \multicolumn{1}{c|}{\multirow{7}{*}{\rotatebox{90} {\textbf{{Reference}}}}} & \multicolumn{1}{c|}{\multirow{7}{*}{\rotatebox{90}{\textbf{{Approach}}}}} & \multicolumn{1}{c|}{\multirow{7}{*}{\rotatebox{90}{\textbf{{Evaluation}}}}} & \multicolumn{4}{c|}{\textbf{{Criteria} }} &
\multicolumn{3}{c|}{\textbf{{Metrics}}} &
\multicolumn{1}{c|}{\multirow{8}{*}{\rotatebox{90} {\textbf{{Category}}}}} &
\multicolumn{1}{c|}{\multirow{7}{*}{\rotatebox{90}{\textbf{{Layer}}}}} &
\multicolumn{1}{c|}{\multirow{7}{*}{\textbf{{Observation}}}} \\ \cline{5-11}

& \multicolumn{1}{c|}{}
& \multicolumn{1}{c|}{}
& \multicolumn{1}{c|}{}
& \multicolumn{1}{c|}{\multirow{7}{*}{\rotatebox{90}{\textbf{{QoS}}}}}
& \multicolumn{1}{c|}{\multirow{6}{*}{\rotatebox{90}{\textbf{{Scalability}}}}}
& \multicolumn{1}{c|}{\multirow{6}{*}{\rotatebox{90}{\textbf{{Heterogeneity}}}}}
& \multicolumn{1}{c|}{\multirow{6}{*}{\rotatebox{90}{\textbf{{Mobility}}}}}
& \multicolumn{1}{c|}{\multirow{6}{*}{\rotatebox{90}{\textbf{{Energy}}}}}
& \multicolumn{1}{c|}{\multirow{6}{*}{\rotatebox{90}{\textbf{{Capacity}}}}}
& \multicolumn{1}{c|}{\multirow{6}{*}{\rotatebox{90}{\textbf{{Latency}}}}}
& \multicolumn{1}{c|}{} & \multicolumn{1}{c|}{} & \multicolumn{1}{c|}{} \\ 

& \multicolumn{1}{c|}{} & \multicolumn{1}{c|}{} & \multicolumn{1}{c|}{} & \multicolumn{1}{c|}{} & \multicolumn{1}{c|}{} & \multicolumn{1}{c|}{} & \multicolumn{1}{c|}{} & \multicolumn{1}{c|}{} & \multicolumn{1}{c|}{} & \multicolumn{1}{c|}{} & \multicolumn{1}{c|}{} & \multicolumn{1}{c|}{} & \multicolumn{1}{c|}{} \\
& \multicolumn{1}{c|}{} & \multicolumn{1}{c|}{} & \multicolumn{1}{c|}{} & \multicolumn{1}{c|}{} & \multicolumn{1}{c|}{} & \multicolumn{1}{c|}{} & \multicolumn{1}{c|}{} & \multicolumn{1}{c|}{} & \multicolumn{1}{c|}{} & \multicolumn{1}{c|}{} & \multicolumn{1}{c|}{} & \multicolumn{1}{c|}{} & \multicolumn{1}{c|}{} \\
& \multicolumn{1}{c|}{} & \multicolumn{1}{c|}{} & \multicolumn{1}{c|}{} & \multicolumn{1}{c|}{} & \multicolumn{1}{c|}{} & \multicolumn{1}{c|}{} & \multicolumn{1}{c|}{} & \multicolumn{1}{c|}{} & \multicolumn{1}{c|}{} & \multicolumn{1}{c|}{} & \multicolumn{1}{c|}{} & \multicolumn{1}{c|}{} & \multicolumn{1}{c|}{} \\
& \multicolumn{1}{c|}{} & \multicolumn{1}{c|}{} & \multicolumn{1}{c|}{} & \multicolumn{1}{c|}{} & \multicolumn{1}{c|}{} & \multicolumn{1}{c|}{} & \multicolumn{1}{c|}{} & \multicolumn{1}{c|}{} & \multicolumn{1}{c|}{} & \multicolumn{1}{c|}{} & \multicolumn{1}{c|}{} & \multicolumn{1}{c|}{} & \multicolumn{1}{c|}{}  \\
& \multicolumn{1}{c|}{} & \multicolumn{1}{c|}{} & \multicolumn{1}{c|}{} & \multicolumn{1}{c|}{} & \multicolumn{1}{c|}{} & \multicolumn{1}{c|}{} & \multicolumn{1}{c|}{} & \multicolumn{1}{c|}{} & \multicolumn{1}{c|}{} & \multicolumn{1}{c|}{} & \multicolumn{1}{c|}{} & \multicolumn{1}{c|}{} & \multicolumn{1}{c|}{}  \\
\hline

\multirow{6}{*}{\rotatebox{90}{Data Center}}
& \cite{eesdn_xiong_2018}
& S 
& Sim, Exp 
& \Circle 
& \CIRCLE 
& \Circle 
& \Circle 
& \CIRCLE 
& \CIRCLE 
& \Circle 
& DA, ML
& C
& {Lightpath management and operations with prediction; Optical; Inter-DC \acs{EON}, BP Neural Network} \\ \cline{2-14}
& \cite{eesdn_xie_2018}  
& H 
& Sim 
& \CIRCLE 
& \CIRCLE
& \LEFTcircle 
& \Circle 
& \CIRCLE 
& \CIRCLE 
& \CIRCLE 
& DA, SM
& C, M
& {Controllers, sw., ports; Multi-controller; TE, domains} \\ \cline{2-14}
& \cite{eesdn_hadi_2019} 
& E, H 
& Sim 
& \CIRCLE 
& \CIRCLE 
& \LEFTcircle 
& \Circle 
& \CIRCLE 
& \CIRCLE 
& \CIRCLE 
& DA, SM
& C
& {Transponder (re)configuration; Optical; Inter-DC EON} \\ \cline{2-14}
& \cite{eesdn_zeng_2019} 
& E, H 
& Sim 
& \Circle 
& \CIRCLE 
& \LEFTcircle 
& \Circle 
& \CIRCLE 
& \Circle 
& \Circle 
& DA, SM
& C
& {Device activation, rule instal., data transmission; \acs{SDDC}} \\ \cline{2-14}
& \cite{eesdn_zhao_2021} 
& S 
& Sim 
& \Circle 
& \CIRCLE 
& \LEFTcircle 
& \Circle 
& \CIRCLE 
& \CIRCLE 
& \Circle 
& DA, SM
& C, M
& {Flow correlation, ALR, device state; SDDC; TE} \\ \cline{2-14}
& \cite{eesdn_zhao_2021_fgcs} 
& S 
& Sim, Exp 
& \Circle 
& \CIRCLE 
& \LEFTcircle 
& \Circle 
& \CIRCLE 
& \CIRCLE 
& \Circle 
& DA, SM
& C, M
& Routing with less device state transition, re-routing to avoid congestion, ALR; SDDC; TE \\ \cline{2-14}
& \cite{eesdn_he_2022} 
& E, H 
& Sim 
& \Circle 
& \CIRCLE 
& \LEFTcircle 
& \CIRCLE 
& \CIRCLE 
& \CIRCLE 
& \Circle 
& DA
& C, M
& Reducing dependencies of multiple migrations; SDDC \\ \cline{2-14}
& \cite{eesdn_canali_2018} 
& H 
& Sim 
&  \Circle 
& \CIRCLE 
& \LEFTcircle 
& \CIRCLE 
& \CIRCLE 
& \CIRCLE 
& \Circle 
& DA
& M
& {Joint computing, data transmission, and migration; TE application; Combined virtual elements and routers} \\ \cline{2-14}
& \cite{eesdn_son_2019}
& H  
& Sim 
& \CIRCLE 
& \Circle 
& \LEFTcircle 
& \Circle 
& \CIRCLE 
& \Circle 
& \CIRCLE 
& DA
& M
& {VM placement near hosts and priority-based bandwidth allocation; TE application} \\ \cline{2-14}
& \cite{eesdn_chaudhary_2020} 
& S 
& Sim 
& \CIRCLE 
& \CIRCLE 
& \Circle 
& \Circle 
& \CIRCLE 
& \CIRCLE 
& \CIRCLE 
& DA, SM
& M
& {Flow re-routing and link utilization; TE application} \\ \cline{2-14}
& \cite{eesdn_wen_2021} 
&  H 
& Sim 
& \Circle 
& \CIRCLE 
& \LEFTcircle 
& \CIRCLE 
& \CIRCLE 
& \CIRCLE 
& \Circle 
& DA, EH
& M
& {Multi-cloud energy consumption and green energy} \\ \hline
\multirow{5}{*}{\rotatebox{90}{Transport}}
& \cite{eesdn_chen_2018} 
& H 
& Sim 
& \CIRCLE 
& \CIRCLE 
& \LEFTcircle 
& \Circle 
& \CIRCLE 
& \CIRCLE 
& \CIRCLE 
& DA
& C
& {Balancing energy and network usage; TE application} \\ \cline{2-14}
& \cite{eesdn_huin_2018} 
& E, H 
& Sim, Exp 
& \CIRCLE 
& \CIRCLE 
& \CIRCLE 
& \Circle 
& \CIRCLE 
& \Circle 
& \CIRCLE 
& DA, SM, HT
& C, M
& {Packet losses when turning off network nodes; Hybrid SDN}; TE application \\ \cline{2-14}
& \cite{eesdn_assefa_2020} 
& E, H 
& Sim, Exp 
& \CIRCLE 
& \CIRCLE 
& \Circle 
& \Circle 
& \CIRCLE 
& \CIRCLE 
& \Circle 
& DA
& C, M
& RESDN metric; TE application and metric \\ \cline{2-14}
& \cite{eesdn_galanjimenez_2020} 
& H 
& Sim 
& \Circle 
& \CIRCLE 
& \Circle 
& \Circle 
& \CIRCLE 
& \CIRCLE 
& \Circle 
& DA, SM
& C
& Reducing TCAM utilization \\ \cline{2-14}
& \cite{eesdn_hiryanto_2021} 
& E, H 
& Sim 
& \CIRCLE 
& \CIRCLE 
& \CIRCLE 
& \Circle 
& \CIRCLE 
& \Circle 
& \CIRCLE 
& DA, SM, HT
& C
& Switching off cables considering bundled links; Hybrid SDN \\ \cline{2-14}
& \cite{eesdn_wang_2022} 
& S 
& Sim 
& \Circle 
& \CIRCLE 
& \Circle 
& \Circle 
& \CIRCLE 
& \Circle 
& \CIRCLE 
& DA
& C, M
& Reducing energy with network topology and traffic adaptation; SDN control on topology and reliability \\ \cline{2-14}
& \cite{eesdn_maity_2022} 
& S 
& Sim 
& \CIRCLE 
& \CIRCLE 
& \LEFTcircle 
& \Circle 
& \CIRCLE 
& \CIRCLE 
& \Circle 
& DA, SM
& C, M
& Switch off SDN controllers and control links; CPP \\ \cline{2-14}
& \cite{eesdn_liu_2018} 
& H 
& Sim 
& \CIRCLE 
& \CIRCLE 
& \LEFTcircle 
&  \Circle 
& \CIRCLE 
& \CIRCLE 
& \Circle 
& DA
& M
& {Model for balancing 3D QoE and energy cost with path selection and rate allocation; Multimedia application} \\ \cline{2-14}
& \cite{eesdn_maaloul_2018} 
& E, H 
& Sim 
&  \Circle 
& \CIRCLE 
& \LEFTcircle 
& \Circle 
& \CIRCLE 
& \CIRCLE 
& \Circle 
& DA, SM
& M
& {Switching off nodes and links considering flow table capacity; TE application} \\ \cline{2-14}
& \cite{eesdn_chen_2020} 
& S 
& Sim 
& \Circle 
& \CIRCLE 
& \Circle 
& \Circle 
& \CIRCLE 
& \CIRCLE 
& \Circle 
& DA, ML
& M
& {Power consumption of switches and ports; Traffic monitoring and prediction; DL} \\ \cline{2-14}
& \cite{eesdn_zhao_2021_systj} 
& S 
& Sim 
& \Circle 
& \CIRCLE 
& \LEFTcircle 
& \Circle 
& \CIRCLE 
& \CIRCLE 
& \Circle 
& DA, SM
& C
& Power-saving and load-balancing trade-off \\ \cline{2-14}
& \cite{eesdn_jimenezlazaro_2022} 
& H 
& Sim 
& \Circle 
& \CIRCLE 
& \Circle 
& \Circle 
& \CIRCLE 
& \CIRCLE 
& \Circle 
& DA, SM, \newline ML
& C
& Genetic algorithm-based with on/off modes and ALR; Prediction \\ \cline{2-14}
& \cite{eesdn_assefa_2021} 
& S
& Sim 
& \Circle
& \CIRCLE 
& \LEFTcircle 
& \Circle
& \CIRCLE 
& \CIRCLE 
& \Circle
& DA, ML
& C
& Utility interval parameters prediction \\ \cline{2-14}
& \cite{eesdn_sacco_2021} 
& S 
& Sim 
& \CIRCLE 
& \CIRCLE 
& \LEFTcircle 
& \Circle 
& \CIRCLE 
& \CIRCLE 
& \Circle 
& DA, SM, \newline ML
& C, M
& Balancing power, QoE, and fairness index; Multi-agent RL \\ \hline
\multirow{5}{*}{\rotatebox{90}{Wireless}}
& \cite{eesdn_abolhasan_2018} 
& S 
& Sim 
& \Circle 
& \CIRCLE 
& \LEFTcircle 
& \CIRCLE 
& \CIRCLE 
& \Circle 
& \Circle 
& DA
& C, M
& {Traffic offloading from cellular to multi-hop D2D networks; TE application} \\  \cline{2-14}
& \cite{eesdn_bera_2019} 
& S 
& Sim 
& \Circle 
& \CIRCLE 
& \LEFTcircle 
& \CIRCLE 
& \CIRCLE 
& \Circle 
& \Circle 
& DA, SM
& C
& Turning off WLAN APs considering traffic with location prediction; WLANs \\ \cline{2-14}
& \cite{eesdn_han_2019} 
& H 
& Sim 
& \Circle 
& \CIRCLE 
& \CIRCLE 
& \Circle 
& \CIRCLE 
& \CIRCLE 
& \Circle 
& DA, HT
& C
& Using MEC to control power and mode operations of base stations; HetNet cellular network \\ \cline{2-14}
& \cite{eesdn_kim_2020} 
& E, H 
& Sim, Exp 
& \CIRCLE 
& \CIRCLE 
& \Circle 
& \CIRCLE 
& \CIRCLE 
& \CIRCLE 
& \Circle 
& DA
& C
& {Interferences and demands of users; WLANs} \\ \cline{2-14}
& \cite{eesdn_yazdinejad_2021} 
& S 
& Sim 
& \CIRCLE 
& \CIRCLE 
& \LEFTcircle 
& \CIRCLE 
& \CIRCLE 
& \CIRCLE 
& \CIRCLE 
& DA
& C
& {Consensus algorithm with controllers; Cellular networks} \\ \cline{2-14}
& \cite{eesdn_luo_2020} 
& S 
& Sim 
& \CIRCLE 
& \CIRCLE 
& \Circle 
& \Circle 
& \CIRCLE 
& \CIRCLE 
& \Circle 
& DA, ML
& M
& Playout buffer, adaptive bitrate, edge caching, video transcoding and transmission; DRL (A3C) \\ \hline
\multirow{5}{*}{\rotatebox{90}{WSN}}
& \cite{eesdn_kumar_2018} 
& H 
& Sim 
& \CIRCLE 
& \CIRCLE 
& \LEFTcircle 
& \Circle 
& \CIRCLE 
& \Circle 
& \CIRCLE 
& DA
& C, M
& Path-loss-based energy model with PSO; TE app. \\ \cline{2-14}
& \cite{eesdn_juradolasso_2020} 
& H 
& Sim 
& \Circle 
& \CIRCLE 
& \Circle 
& \Circle 
& \CIRCLE 
& \CIRCLE 
& \Circle 
& DA
& C
& Neighbor discovery and advertisement, configuration, control, transmission; Topology control \\ \cline{2-14}
& \cite{eesdn_chaudhry_2022} 
& H 
& Sim 
& \CIRCLE 
& \CIRCLE 
& \LEFTcircle 
& \Circle 
& \CIRCLE 
& \Circle 
& \CIRCLE 
& DA
& C, M
& Multi-objective considering energy, latency, and load balancing; Control node placement \\ \cline{2-14}
& \cite{eesdn_fang_2022} 
& S 
& Sim 
& \CIRCLE 
& \CIRCLE 
& \Circle 
& \Circle 
& \CIRCLE 
& \Circle 
& \CIRCLE 
& DA
& C, M
& ECPUB metric with load balancing; TE/Security app. \\ \hline
\end{tabular}
\end{table*}
	
\begin{table*}[t]
\ContinuedFloat
\centering
\caption{Survey on Energy-Efficient SDN (Contd.) \\ (\Circle Not considered \LEFTcircle Partially considered \CIRCLE Considered)}
\begin{tabular}{|c|p{0.55cm}|p{0.6cm}|p{0.5cm}|p{0.2cm}|p{0.2cm}|p{0.2cm}|p{0.2cm}|p{0.2cm}|p{0.2cm}|p{0.2cm}|p{1.1cm}|p{0.6cm}|p{6.8cm}|}
\hline

& \cite{eesdn_kumar_2022} 
& E, S 
& Sim 
& \CIRCLE 
& \CIRCLE 
& \LEFTcircle 
& \Circle 
& \CIRCLE 
& \Circle 
& \CIRCLE 
& SM
& C, M
& Processing tasks, memory operations, neighbour sensing \\ \cline{2-14}
& \cite{eesdn_younus_2022} 
& S 
& Sim, Exp 
& \CIRCLE 
& \CIRCLE 
& \Circle 
& \Circle 
& \CIRCLE 
& \Circle 
& \Circle 
& DA, ML
& C, M
& Distance, free space, multiple paths; RL \\ \hline
\multirow{6}{*}{\rotatebox{90}{Emerging}}
& \cite{eesdn_yazdinejad_2020} 
& S 
& Exp 
& \CIRCLE 
& \CIRCLE 
& \LEFTcircle 
& \Circle 
& \CIRCLE 
& \CIRCLE 
& \CIRCLE 
& DA
& C
& {Cluster architecture using blockchain; IoT} \\ \cline{2-14}
& \cite{eesdn_tu_2020} 
& S 
& Sim 
& \Circle 
& \CIRCLE 
& \LEFTcircle 
& \Circle 
& \CIRCLE 
& \Circle 
& \Circle 
& DA
& C, M
& {Inter-satellite and link switching, Security app.; Satellite networks} \\ \cline{2-14}
& \cite{eesdn_yang_2021} 
& S, P 
& Sim 
& \CIRCLE 
& \CIRCLE 
& \Circle 
& \Circle 
& \CIRCLE 
& \Circle 
& \CIRCLE 
& DA, SM
& C
& Recalculation of channel path-loss model; Maritime networks \\ \cline{2-14}
& \cite{eesdn_li_2022} 
& S 
& Sim  
& \CIRCLE 
& \CIRCLE 
& \Circle 
& \CIRCLE 
& \CIRCLE 
& \Circle 
& \Circle 
& DA
& C
& User association, location, BS RA, and data; UAVs \\ \cline{2-14}
& \cite{eesdn_jiang_2021} 
& H 
& Sim 
& \Circle 
& \CIRCLE 
& \CIRCLE 
& \CIRCLE 
& \CIRCLE 
& \CIRCLE 
& \Circle 
& DA, HT
& C, M
& Energy transmission, port, device, (de)multiplexing; Vehicular networks \\ \cline{2-14}
& \cite{eesdn_boukerche_2022} 
& S 
& Sim 
& \CIRCLE 
& \CIRCLE 
& \LEFTcircle 
& \CIRCLE 
& \CIRCLE 
& \CIRCLE 
& \CIRCLE 
& DA
& C
& Clustering with split-and-merge; Vehicular networks \\ \cline{2-14}
& \cite{eesdn_liu_2022} 
& S 
& Sim 
& \CIRCLE 
& \CIRCLE 
& \CIRCLE 
& \Circle 
& \CIRCLE 
& \CIRCLE 
& \CIRCLE 
& DA, HT, \newline ML
& C, M
& Different computing powers and latency; Vehicular networks; RL \\ \cline{2-14}
& \cite{eesdn_aujla_2019} 
& S 
& Sim 
& \CIRCLE 
& \CIRCLE 
& \LEFTcircle 
& \CIRCLE 
& \CIRCLE 
& \CIRCLE 
& \CIRCLE 
& DA, SM, \newline EH
& C, M
& Servers, devices, cooling equipment, facilities, renewable energy; EDC, vehicular networks \\ \cline{2-14}
& \cite{eesdn_cicioglu_2020} 
& S 
& Sim 
& \CIRCLE 
& \CIRCLE 
& \LEFTcircle 
& \Circle 
& \CIRCLE 
& \CIRCLE 
& \CIRCLE 
& DA
& M
& Fuzzy Djikstra-based routing; TE application; WBANs \\ \cline{2-14}
& \cite{eesdn_ruby_2021} 
& S 
& Sim 
& \CIRCLE 
& \CIRCLE 
& \LEFTcircle 
& \Circle 
& \CIRCLE 
& \Circle 
& \CIRCLE 
& DA
& M
& Interference of ad hoc scenarios; TE application, Multi-modal UWSNs \\
\hline
\end{tabular}
\begin{tablenotes}
\item E: Exact, H: Heuristic, S: Scheme; Sim: Simulation, Exp: Experimentation; 	DA: Dynamic Adaptation, SM: Sleep Modes, HT: Heterogeneous Resources, EH: Energy Harvesting, ML: Machine Learning; C: Control Layer, M: Management Layer
\end{tablenotes}
\end{table*}

\subsubsection{Data center networks}
In this network scenario with the control layer as the research focus, SDN is used for handling intra-DC and inter-DC connections, multi-controller, flow management, and VM migrations to assist computing tasks. Xiong et al. \cite{eesdn_xiong_2018} proposed centralized management of lightpath in EON with the use of traffic prediction---leveraging back propagation (BP), neural networks (NN), and particle swarm optimization (PSO)---to minimizing termination and re-setup operations of lightpath that affect power consumption. In similar case, Hadi et al. \cite{eesdn_hadi_2019} examined transponder configuration problem with constraints of QoS and physical requirements of OFDM-based inter-DC EON. Xie et al. \cite{eesdn_xie_2018} argued that distributed power control is needed in large scale SDDC considering flow routing in the intra- and inter-domain, as well as energy efficiency in the control plane with a distributed manner. Zeng et al. \cite{eesdn_zeng_2019} considered joint device activation, rule installation, and data transmission for reducing energy consumption. Zhao et al. \cite{eesdn_zhao_2021} constructed a framework to jointly study power efficiency, link congestion, and device state transition. Similar to the aforementioned authors, Zhao et al. \cite{eesdn_zhao_2021_fgcs} examined device state transition by reducing the changes so that it will not consume substantial energy. 

Traffic engineering (TE) applications were discussed in several papers. Xie et al. \cite{eesdn_xie_2018} proposed distributed flow routing in intra- and inter-domain, considering energy efficiency at the control plane. Zhao et al. \cite{eesdn_zhao_2021} focused on a framework for optimizing power, considering the efficiency, link congestion, and state transition of network devices. Zhao et al. \cite{eesdn_zhao_2021_fgcs} studied device state transition by reducing it so that sizable power, QoS, etc. can be reduced. He et al. \cite{eesdn_he_2022} discussed on multiple concurrent VMs migration for minimizing interference of migration and making short time convergence in cloud DCs. Canali et al. \cite{eesdn_canali_2018} investigated joint energy cost problem of computing, data transmission, and migration in SDDC. Son et al. \cite{eesdn_son_2019} proposed algorithms based on policies to allocate VMs and their respective bandwidth requirements. Chaudhary et al. \cite{eesdn_chaudhary_2020} discussed energy-aware routing problem in DC with policies to drive scheduling, routing, and re-routing. Finally, Wen et al. \cite{eesdn_wen_2021} worked on workload scheduling in multi-cloud connected with SD-WAN and the use of conventional and renewable energy.

\subsubsection{Transport networks}
Chen et al. \cite{eesdn_chen_2018} proposed allocation of caching, computing, bandwidth jointly to build a framework to orchestrate the three resources. Huin et al. \cite{eesdn_huin_2018} discussed hybrid SDN with optimization of energy to cater the need of energy-aware routing by setting to sleep SDN devices and turning off links. They conducted the research by simulation and SDN testbed. Assefa and Ozkazap \cite{eesdn_assefa_2020} proposed RESDN metric as a ratio to measure energy efficiency with the intervals of link utility in the network. Hiryanto et al. \cite{eesdn_hiryanto_2021} examined the problem of minimizing energy consumption by using multiple stages of switch upgrading with available budget in every stage. In this problem, the authors use bundled links that can be switched off, upgrade and traffic considerations, and alternative path. Wang et al. \cite{eesdn_wang_2022} discussed about the problem of topology switching along with the reliability. With this dynamic topology, the energy consumption will be reduced. Maity et al. \cite{eesdn_maity_2022} argued that by partitioning the network into several partitions with turning off links but still connected of the controller, the energy consumption will be reduced. There are two network applications that were found in the classified references: TE and multimedia. The earlier case were combined with control layer in several articles \cite{eesdn_huin_2018, eesdn_assefa_2020, eesdn_wang_2022, eesdn_maity_2022}. For the latter, the costs of energy are balanced with 3D QoE with path selection and allocation of data rate by Liu et al. \cite{eesdn_liu_2018}, flow table capacity in SDN switch consideration to turn off the devices and links by Maaloul et al. \cite{eesdn_maaloul_2018}. In transport networks scenario, the consideration of QoS is used in several papers along with metric related to utilization of the links. Hybrid SDN with supports of technology---optical networks in particular---need to be enabled by the exposure of the features of the network devices. Network topology as well as traffic adaptation using DA strategy can be utilized in this scenario due to the need of the interconnections to multiple high-speed devices.

\subsubsection{Wireless networks}
Abolhasan et al. \cite{eesdn_abolhasan_2018} discussed mobile or cellular networks scenarios. They tried to reduce traffic overhead in LTE networks with SDN controller resulting higher scalability. Bera et al. \cite{eesdn_bera_2019} gave a scheme for multi-controller placement with a resource management model considering latency. WSNs in this network scenario covered most of the collected literature. Han et al. \cite{eesdn_han_2019} integrated SDN into WSNs, considering residual energy, transmission power, and game theory for extending network lifetime. Kim et al. \cite{eesdn_kim_2020} implemented SDN in WSN with the goal of performance assessment, including energy consumption. Yazdinejad et al. \cite{eesdn_yazdinejad_2021} and Luo et al. \cite{eesdn_luo_2020} proposed a heuristic for allocating energy with energy averaging and minimization problem by choosing relay nodes and de-duplicated data flows.

\subsubsection{WSNs}
The \acs{RL} approach was used by Younus et al. \cite{eesdn_younus_2022} to optimize routing in software-defined WSN (SDWSN) by proposing a controller framework which could generate routing tables and provide better network lifetime compared to non-SDWSN. Kumar et al. \cite{eesdn_kumar_2018}. Jurado Lasso et al. \cite{eesdn_juradolasso_2020}. Chaudhry et al. \cite{eesdn_chaudhry_2022}. Fang et al. \cite{eesdn_fang_2022}. Finally, Kumar et al. \cite{eesdn_kumar_2022}.

\subsubsection{Emerging networks}
There are five categories in this network scenario: IoT networks, satellite networks, WBAN, vehicular networks, and mobile networks in MEC environment. Yazdinejad et al. \cite{eesdn_yazdinejad_2020} discussed using SDN in IoT networks utilizing blockchain. They focused on secure SDN controller architecture with a scheme using cluster structure. Energy efficiency in satellite networks was discussed in \cite{eesdn_tu_2020}, with consideration of link switching and inter-satellite link energy consumption. Yang et al. \cite{eesdn_yang_2021} proposed a scheme with multi-objective spider monkey optimization to reduce congestion and thermal effects. Li et al. \cite{eesdn_li_2022} and Jiang et al. \cite{eesdn_jiang_2021} followed up discussions in distributing resource and communications optimization in vehicular fog environments, and energy and \acs{QoE} optimization with video streaming applications in mobile networks with MEC. Both studies used RF approach to propose the solutions.

\subsection{Energy-Efficient NFV}
In this section we discuss the survey on energy-efficient NFV. After classified by network scenario, energy-efficiency strategies, and control/MANO layer, each study was detailed in respect to its attributes, which were defined in Section \ref{section:survey_ee_netsoft}, and compared the results in Table \ref{table:es-nfv}.

\begin{table*}
\centering
\caption{Survey on Energy-Efficient NFV\\ (\Circle Not considered \LEFTcircle Partially considered \CIRCLE Considered)}
\label{table:es-nfv}
\begin{tabular}{|c|p{0.55cm}|p{0.6cm}|p{0.5cm}|p{0.2cm}|p{0.2cm}|p{0.2cm}|p{0.2cm}|p{0.2cm}|p{0.2cm}|p{0.2cm}|p{1cm}|p{0.4cm}|p{7cm}|}
	\hline
	\multirow{7}{*}{\rotatebox{90}{\textbf{Netw. Scenario}}} & \multicolumn{1}{c|}{\multirow{7}{*}{\rotatebox{90} {\textbf{{Reference}}}}} & \multicolumn{1}{c|}{\multirow{7}{*}{\rotatebox{90}{\textbf{{Approach}}}}} & \multicolumn{1}{c|}{\multirow{7}{*}{\rotatebox{90}{\textbf{{Evaluation}}}}} & \multicolumn{4}{c|}{\textbf{{Criteria} }} &
	\multicolumn{3}{c|}{\textbf{{Metrics}}} &
	\multicolumn{1}{c|}{\multirow{8}{*}{\rotatebox{90} {\textbf{{Category}}}}} &
	\multicolumn{1}{c|}{\multirow{7}{*}{\rotatebox{90}{\textbf{{Layer}}}}} &
	\multicolumn{1}{c|}{\multirow{7}{*}{\textbf{{Observation}}}} \\ \cline{5-11}

& \multicolumn{1}{c|}{}
& \multicolumn{1}{c|}{}
& \multicolumn{1}{c|}{}
& \multicolumn{1}{c|}{\multirow{7}{*}{\rotatebox{90}{\textbf{{QoS}}}}}
& \multicolumn{1}{c|}{\multirow{6}{*}{\rotatebox{90}{\textbf{{Scalability}}}}}
& \multicolumn{1}{c|}{\multirow{6}{*}{\rotatebox{90}{\textbf{{Heterogeneity}}}}}
& \multicolumn{1}{c|}{\multirow{6}{*}{\rotatebox{90}{\textbf{{Mobility}}}}}
& \multicolumn{1}{c|}{\multirow{6}{*}{\rotatebox{90}{\textbf{{ Energy}}}}}
& \multicolumn{1}{c|}{\multirow{6}{*}{\rotatebox{90}{\textbf{{Capacity}}}}}
& \multicolumn{1}{c|}{\multirow{6}{*}{\rotatebox{90}{\textbf{{Latency}}}}}
& \multicolumn{1}{c|}{} & \multicolumn{1}{c|}{} & \multicolumn{1}{c|}{} \\ 

& \multicolumn{1}{c|}{} & \multicolumn{1}{c|}{} & \multicolumn{1}{c|}{} & \multicolumn{1}{c|}{} & \multicolumn{1}{c|}{} & \multicolumn{1}{c|}{} & \multicolumn{1}{c|}{} & \multicolumn{1}{c|}{} & \multicolumn{1}{c|}{} & \multicolumn{1}{c|}{} & \multicolumn{1}{c|}{} & \multicolumn{1}{c|}{} & \multicolumn{1}{c|}{} \\
& \multicolumn{1}{c|}{} & \multicolumn{1}{c|}{} & \multicolumn{1}{c|}{} & \multicolumn{1}{c|}{} & \multicolumn{1}{c|}{} & \multicolumn{1}{c|}{} & \multicolumn{1}{c|}{} & \multicolumn{1}{c|}{} & \multicolumn{1}{c|}{} & \multicolumn{1}{c|}{} & \multicolumn{1}{c|}{} & \multicolumn{1}{c|}{} & \multicolumn{1}{c|}{} \\
& \multicolumn{1}{c|}{} & \multicolumn{1}{c|}{} & \multicolumn{1}{c|}{} & \multicolumn{1}{c|}{} & \multicolumn{1}{c|}{} & \multicolumn{1}{c|}{} & \multicolumn{1}{c|}{} & \multicolumn{1}{c|}{} & \multicolumn{1}{c|}{} & \multicolumn{1}{c|}{} & \multicolumn{1}{c|}{} & \multicolumn{1}{c|}{} & \multicolumn{1}{c|}{} \\
& \multicolumn{1}{c|}{} & \multicolumn{1}{c|}{} & \multicolumn{1}{c|}{} & \multicolumn{1}{c|}{} & \multicolumn{1}{c|}{} & \multicolumn{1}{c|}{} & \multicolumn{1}{c|}{} & \multicolumn{1}{c|}{} & \multicolumn{1}{c|}{} & \multicolumn{1}{c|}{} & \multicolumn{1}{c|}{} & \multicolumn{1}{c|}{} & \multicolumn{1}{c|}{}  \\
& \multicolumn{1}{c|}{} & \multicolumn{1}{c|}{} & \multicolumn{1}{c|}{} & \multicolumn{1}{c|}{} & \multicolumn{1}{c|}{} & \multicolumn{1}{c|}{} & \multicolumn{1}{c|}{} & \multicolumn{1}{c|}{} & \multicolumn{1}{c|}{} & \multicolumn{1}{c|}{} & \multicolumn{1}{c|}{} & \multicolumn{1}{c|}{} & \multicolumn{1}{c|}{}  \\
\hline

\multirow{6}{*}{\rotatebox{90}{Data Center}}
& \cite{eenfv_assi_2019} 
& H 
& Sim 
& \Circle 
& \CIRCLE 
& \LEFTcircle 
& \Circle 
& \CIRCLE 
& \CIRCLE 
& \Circle 
& DA, SM
& RO
& {Flow scheduling for different VNFs; VNF-SCH} \\ \cline{2-14}
& \cite{eenfv_bolla_2020} 
& E, S 
& Sim, Exp 
& \CIRCLE 
& \CIRCLE 
& \LEFTcircle 
& \Circle 
& \CIRCLE 
& \CIRCLE 
& \CIRCLE 
& DA, SM
& RO
& {VNF workload profiling and KPIs or metrics estimation; VNF-DPL} \\ \cline{2-14}
& \cite{eenfv_padhy_2021} 
& E, H 
& Sim 
& \Circle 
& \CIRCLE 
& \LEFTcircle 
& \CIRCLE 
& \CIRCLE 
& \CIRCLE 
& \Circle 
& DA, SM
& RO
& {VNF scaling and migration; VNF-RCF} \\ \cline{2-14}
& \cite{eenfv_ortin_2022} 
& H 
& Sim 
& \Circle 
& \CIRCLE 
& \LEFTcircle 
& \Circle 
& \CIRCLE 
& \CIRCLE 
& \Circle 
& DA, SM
& RO
& {Activation, service lifetime, failure probability; VNF-RCF} \\ \cline{2-14}
& \cite{eenfv_chaurasia_2022} 
& S 
& Exp 
& \Circle 
& \CIRCLE
& \CIRCLE 
& \Circle 
& \CIRCLE 
& \CIRCLE 
& \Circle 
& DA, HT
& RO
& {Using GPU cores and CPU-GPU sync.; VNF-DPL, VNF-TR} \\ \cline{2-14}
& \cite{eenfv_taskou_2021} 
&  E, H 
&  Sim 
&  \CIRCLE 
&  \CIRCLE 
&  \Circle 
&  \Circle 
&  \CIRCLE 
&  \CIRCLE 
&  \CIRCLE 
&  DA, SM
&  SO
& {Multi-objective with resource util. cost; VNF-P, VNF-TR} \\ \cline{2-14}
& \cite{eenfv_abdelaal_2021} 
& H 
& Sim 
& \Circle 
& \CIRCLE 
& \LEFTcircle 
& \CIRCLE 
& \CIRCLE 
& \CIRCLE 
& \Circle 
& DA, SM
& SO
& {Service availability and reconfig. with migration; VNF-RCF} \\ \hline
\multirow{5}{*}{\rotatebox{90}{Transport}}
& \cite{eenfv_kaur_2019} 
& H 
& Sim 
& \Circle 
& \CIRCLE 
& \Circle 
& \Circle 
& \CIRCLE 
& \CIRCLE 
& \Circle 
& DA
& RO
& {VNF deployment over multi-domain networks; VNF-DPL} \\ \cline{2-14}
& \cite{eenfv_montazerolghaem_2020} 
& E, S 
& Sim, Exp 
& \CIRCLE 
& \CIRCLE 
& \LEFTcircle 
& \CIRCLE 
& \CIRCLE 
& \CIRCLE 
& \CIRCLE 
& DA, SM
& RO, SO
& {Load balancing using processor, memory, and temperature monitoring; VNF-DPL, VNF-TR} \\ \cline{2-14}
& \cite{eenfv_demirci_2021} 
& E, H 
& Sim 
& \Circle 
& \CIRCLE 
& \LEFTcircle 
& \Circle 
& \CIRCLE 
& \Circle 
& \Circle 
& DA, SM
& RO
& {Flow-based security reqs. and resources; VNF-P, VNF-DPL} \\ \cline{2-14}
& \cite{eenfv_chen_2021} 
& S 
& Sim 
& \Circle 
& \CIRCLE 
& \Circle 
& \Circle 
& \CIRCLE 
& \CIRCLE 
& \Circle 
& DA, SM
& RO
& {Network load and subst. network's scale adjustment; VNF-P} \\ \cline{2-14}
& \cite{eenfv_zhang_2019} 
& E, H 
& Sim  
& \Circle 
& \CIRCLE 
& \LEFTcircle 
& \CIRCLE 
& \CIRCLE 
& \CIRCLE 
& \Circle 
& DA, SM
& SO
& {Power consump. of servers and phy. links; VNF-P, VNF-TR} \\ \cline{2-14}
& \cite{eenfv_zhang_2020} 
&  H 
&  Sim 
&  \Circle 
&  \CIRCLE 
&  \LEFTcircle 
&  \Circle 
&  \CIRCLE 
&  \CIRCLE 
&  \Circle 
&  DA, SM
&  SO
& {Load balancing in multi-domain SDNs; VNF-P, VNF-TR} \\ \cline{2-14}
& \cite{eenfv_mai_2021} 
& H 
& Sim 
& \Circle 
& \CIRCLE 
& \Circle 
& \Circle 
& \CIRCLE 
& \Circle 
& \Circle 
& DA, SM
& SO
& {Service availability reqs. with backup paths; VNF-P, VNF-TR} \\ \hline
\multirow{5}{*}{\rotatebox{90}{Wireless}}
& \cite{eenfv_al-quzweeni_2019} 
& E, H 
& Sim 
& \Circle 
& \CIRCLE 
& \LEFTcircle 
& \Circle 
& \CIRCLE 
& \CIRCLE 
& \Circle 
& DA, SM
& RO
& {Number of VMs, their locations, util. of servers; VNF-P} \\ \cline{2-14}
& \cite{eenfv_feng_2019} 
& H 
& Sim 
& \Circle 
& \CIRCLE 
& \LEFTcircle 
& \Circle 
& \CIRCLE 
& \Circle 
& \Circle 
& DA
& RO, SO
& {Activation of protection and recovery functions, jointly optimized with traffic routing; VNF-P, VNF-TR} \\ \cline{2-14}
& \cite{eenfv_arzo_2020} 
& E 
& Sim 
& \CIRCLE 
& \CIRCLE 
& \LEFTcircle 
& \Circle 
& \CIRCLE 
& \CIRCLE 
& \CIRCLE 
& DA
& RO
& {Service diff., backup VNFs, CPU overprov., E2E; VNF-P} \\ \cline{2-14}
& \cite{eenfv_tao_2021} 
& H 
& Sim 
& \CIRCLE 
& \CIRCLE 
& \Circle 
& \Circle 
& \CIRCLE 
& \Circle
& \CIRCLE 
& DA
& RO
& {Latency and distance to edge systems; VNF-P} \\ \cline{2-14}
& \cite{eenfv_ismail_2020} 
& H 
& Sim 
& \Circle 
& \CIRCLE 
& \Circle 
& \Circle 
& \CIRCLE 
& \CIRCLE 
& \Circle 
& SM
& RO
& {Functional splits with remote sites; VNF-DPL} \\ \cline{2-14}
& \cite{eenfv_gholipoor_2021} 
& S 
& Sim 
& \CIRCLE 
& \CIRCLE 
& \LEFTcircle 
& \Circle 
& \CIRCLE 
& \CIRCLE 
& \CIRCLE 
& SM, ML
& RO, SO
& {Power and spectrum resources in radio segment, QoS in core segment, E2E; VNF-P, VNF-SCH, VNF-TR} \\ \hline
\multirow{5}{*}{\rotatebox{90}{Emerging}}
& \cite{eenfv_wu_2019} 
& S 
& Sim, Exp 
& \Circle 
& \CIRCLE 
& \LEFTcircle 
& \CIRCLE 
& \CIRCLE 
& \Circle 
& \Circle 
& SM
& RO
& {Topology control, sleep mode, sensor lifetime; VNF-DPL; CPS: WSN, IoT} \\ \cline{2-14}
& \cite{eenfv_chaudhry_2020} 
& S 
& Exp 
& \CIRCLE 
& \CIRCLE 
& \CIRCLE 
& \Circle 
& \CIRCLE 
& \CIRCLE 
& \CIRCLE 
& DA, HT
& RO
& {Microservices, FPGA-based accelerator; VNF-DPL; \acs{EC}} \\ \cline{2-14}
& \cite{eenfv_xu_2020} 
& E, H 
& Sim 
& \CIRCLE 
& \CIRCLE 
& \CIRCLE 
& \CIRCLE 
& \CIRCLE 
& \CIRCLE 
& \CIRCLE 
& DA, HT
& RO, SO
& {Multi-tier cloud and edge clouds, IoT devices' mobility and their energy statuses; VNF-P, VNF-TR; EC} \\ \cline{2-14}
& \cite{eenfv_zhou_2020} 
& S 
& Sim 
& \CIRCLE 
& \CIRCLE 
& \LEFTcircle 
& \Circle 
& \CIRCLE 
& \CIRCLE 
& \CIRCLE 
& DA
& RO
& {Comput., caching, commun. with latency constraint; EC} \\ \cline{2-14}
& \cite{eenfv_tinini_2019} 
& E, H 
& Sim 
& \CIRCLE 
& \CIRCLE 
& \CIRCLE 
& \Circle 
& \CIRCLE 
& \CIRCLE 
& \CIRCLE 
& DA, HT
& RO
& {Wavelength dimensioning; VNF-P; Cloud-Fog RAN} \\ \cline{2-14}
& \cite{eenfv_tipantuna_2021} 
& E, H 
& Sim 
& \Circle 
& \CIRCLE 
& \LEFTcircle 
& \Circle 
& \CIRCLE 
& \CIRCLE 
& \Circle 
& DA, EH
& RO
& {Adapting demands to supplies; VNF-DPL; \acs{EM}, IoT} \\ \cline{2-14}
& \cite{eenfv_zhou_2022} 
& S 
& Sim 
& \Circle 
& \CIRCLE 
& \CIRCLE 
& \Circle 
& \CIRCLE 
& \CIRCLE 
& \Circle 
& DA, HT, ML
& RO, SO
& {QoS differentation with multi-mode green (battery-powered) IoT devices, flow scheduling; VNF-P, VNF-TR; RL, IoT} \\ \cline{2-14}
& \cite{eenfv_gao_2022} 
& S 
& Sim 
& \CIRCLE 
& \CIRCLE 
& \LEFTcircle 
& \Circle 
& \CIRCLE 
& \CIRCLE 
& \CIRCLE 
& SM
& RO, SO
& {Running states of edge servers on satellite nodes, competing user requests; VNF-P, VNF-DPL; Satellite EC} \\ \cline{2-14}
& \cite{eenfv_jia_2021} 
& E, H 
& Sim 
& \Circle 
& \CIRCLE 
& \Circle 
& \CIRCLE
& \CIRCLE 
& \CIRCLE 
& \Circle 
& DA
& SO
& {Commun., energy and comput. resources; VNF-TR; Satellite} \\ \cline{2-14}
& \cite{eenfv_tipantuna_2019} 
& S 
& Sim, Exp 
& \Circle 
& \CIRCLE 
& \Circle 
& \CIRCLE 
& \CIRCLE 
& \CIRCLE 
& \Circle 
& DA
& SO
& {Execution and replacement of services with availability levels and specific time intervals; VNF-SCH; UAV} \\ \cline{2-14}
& \cite{eenfv_pourghasemian_2022} 
& S 
& Sim 
& \CIRCLE 
& \CIRCLE 
& \LEFTcircle 
& \CIRCLE 
& \CIRCLE 
& \CIRCLE 
& \CIRCLE 
& DA, ML
& SO
& {Trajectory design; VNF-RCF; DRL, UAV} \\ \hline
\end{tabular}
\begin{tablenotes}
	\item E: Exact, H: Heuristic, S: Scheme, Sim: Simulation, Exp: Experimentation; 	DA: Dynamic Adaptation, SM: Sleep Modes, HT: Heterogeneous Resources, EH: Energy Harvesting; RO: Resource Orchestration, SO: Service Orchestration
\end{tablenotes}
\end{table*}

\subsubsection{Data center networks}
In this network scenario, the most discussed topic was VNF deployment. Padhy et al. \cite{eenfv_padhy_2021} gave techniques for mobile networks, in terms of exact and heuristic approaches for virtualized mobile core NFs, considering total number of active users, backhaul/fronthaul configuration, and VM inter-traffic. Service deployment problem was the focus of Ortin et al. \cite{eenfv_ortin_2022}, with considerations of limited traffic processing capacity of VNF instances, and management concerns. Taskou et al. \cite{eenfv_taskou_2021} covered serverless computing for merging MEC and NFV at the system level, and deploying VNFs on demand by combining MEC functional blocks with an NFV orchestrator. Assi et al. \cite{eenfv_assi_2019} proposed VNF placement with focus on the problem of assigning and scheduling flows to VNFs that are already placed in physical machines, and traffic flows with deadlines. A VNF consolidation technique was proposed in \cite{eenfv_chaurasia_2022} to determine appropriate servers to be turned off to reduce energy consumption, bandwidth usage, and migration costs. Bolla et al. \cite{eenfv_bolla_2020} proposed a model-based analytics approach for profiling VNF workloads to estimate network key performance indicators (KPIs), power, and latency.

\subsubsection{Transport networks}
The VNF placement problem was discussed in three papers. Zhang et al. \cite{eenfv_zhang_2019} proposed joint optimization, considering locations and traffic steering, while minimizing energy consumption. Kaur et al. \cite{eenfv_kaur_2019} provided placement and chaining in a metro wavelength-division multiplexing (WDM) network with MEC resources, considering virtual topology design on top of the underlying optical backhaul network. Chen et al. \cite{eenfv_chen_2021} focused on optimization considering resource utilization with binary integer programming and a heuristic algorithm. Studies in \cite{eenfv_demirci_2021, eenfv_zhang_2020} targeted SFC with energy efficiency objective in telecom networks. The first article discussed NFV coupled with SDN considering SFC. The second article proposed a resource allocation architecture that enabled energy-aware SFC for SDNs with delay, link and server utilization constraints. The third article focused on SFC on split paths to use fragmented resources and minimize the number of active nodes. The fourth article gave a solution in the form of SFC orchestration targeting reducing carbon footprints using more energy at locations with surplus renewable energy. The fifth article provided NFV resource allocation to users’ SFCs without an orchestrator using blockchain. It is not only minimized energy consumption but also utilized resource cost. Combination NFV with SDN was discussed in \cite{eenfv_montazerolghaem_2020}, with a multimedia application for preventing overload and green networking supports in VoIP centers. A fog architecture was composed using high-capacity and low-capacity fog nodes residing near terminals by Mai et al. \cite{eenfv_mai_2021}.

\subsubsection{Wireless networks}
There are two topics surrounding this scenario, based on the survey. The first one was NFV configuration and orchestration with Wu et al. \cite{eenfv_wu_2019} proposing a systematic virtual networking architecture to perform global virtualization control and monitoring of a \acs{CPS} with NFV. The second one was VNF deployment, with three papers discussing a hybrid cloud-fog RAN architecture that consists of fog computing and NFV to replicate the processing capacity of a CRAN in local fog nodes closed to the remote radio heads \cite{eenfv_tinini_2019}, the joint problem of security function activation and traffic routing in a wireless multi-hop network \cite{eenfv_feng_2019}, and functional splits with NFV in the dual-site virtualized RAN \cite{eenfv_ismail_2020}. Even though generally VNF placement was the most discussed topic in NFV, only one study was found that investigated optimal VNF placement with service differentiation considering six 5G parameters \cite{eenfv_arzo_2020}.

\subsubsection{Emerging networks}
The first study in emerging networks scenarios related to VNF deployment was by Chaudhry et al. \cite{eenfv_chaudhry_2020}, who proposed a joint optimization of resource allocation and \acs{UAV} trajectory in the 3D spaces. Tipantuna et al. \cite{eenfv_tipantuna_2019} discussed a long-term sustainable demand-response architecture using NFV/SDN for efficient management in IoT infrastructure. Siasi et al.  investigated a fog architecture, considering heterogeneity consisting low and high capacity for servers in locations near end devices. This study used \acs{DL} for provisioning SFC in the architecture.

\subsection{Energy-Efficient NS}
In this section, we discuss the survey on energy saving with NS. Based on network scenarios, we detailed each paper according to the previously defined attributes. The results are compared in Table \ref{table:es-ns}.

\acs{VNE} in this survey was researched in three domains: telecom, data center, and enterprise networks. In other scenarios, the investigations examined NS with specific requirements to a scenario as well as emerging trends, such as multi-tenant, MEC, and IoT.

\begin{table*}
\centering
\caption{Survey on Energy-Efficient NS \\ (\Circle Not considered \LEFTcircle Partially considered \CIRCLE Considered)}
\label{table:es-ns}
\begin{tabular}{|c|p{0.55cm}|p{0.6cm}|p{0.5cm}|p{0.2cm}|p{0.2cm}|p{0.2cm}|p{0.2cm}|p{0.2cm}|p{0.2cm}|p{0.2cm}|p{1.1cm}|p{0.4cm}|p{7cm}|}
\hline
\multirow{7}{*}{\rotatebox{90}{\textbf{Netw. Scenario}}} & \multicolumn{1}{c|}{\multirow{7}{*}{\rotatebox{90} {\textbf{{Reference}}}}} & \multicolumn{1}{c|}{\multirow{7}{*}{\rotatebox{90}{\textbf{{Approach}}}}} & \multicolumn{1}{c|}{\multirow{7}{*}{\rotatebox{90}{\textbf{{Evaluation}}}}} & \multicolumn{4}{c|}{\textbf{{Criteria} }} &
\multicolumn{3}{c|}{\textbf{{Metrics}}} &
\multicolumn{1}{c|}{\multirow{8}{*}{\rotatebox{90} {\textbf{{Category}}}}} &
\multicolumn{1}{c|}{\multirow{7}{*}{\rotatebox{90}{\textbf{{Layer}}}}} &
\multicolumn{1}{c|}{\multirow{7}{*}{\textbf{{Observation}}}} \\ \cline{5-11}
	
& \multicolumn{1}{c|}{}
& \multicolumn{1}{c|}{}
& \multicolumn{1}{c|}{}
& \multicolumn{1}{c|}{\multirow{7}{*}{\rotatebox{90}{\textbf{{QoS}}}}}
& \multicolumn{1}{c|}{\multirow{6}{*}{\rotatebox{90}{\textbf{{Scalability}}}}}
& \multicolumn{1}{c|}{\multirow{6}{*}{\rotatebox{90}{\textbf{{Heterogeneity}}}}}
& \multicolumn{1}{c|}{\multirow{6}{*}{\rotatebox{90}{\textbf{{Mobility}}}}}
& \multicolumn{1}{c|}{\multirow{6}{*}{\rotatebox{90}{\textbf{{ Energy}}}}}
& \multicolumn{1}{c|}{\multirow{6}{*}{\rotatebox{90}{\textbf{{Capacity}}}}}
& \multicolumn{1}{c|}{\multirow{6}{*}{\rotatebox{90}{\textbf{{Latency}}}}}
& \multicolumn{1}{c|}{} & \multicolumn{1}{c|}{} & \multicolumn{1}{c|}{} \\ 
	
& \multicolumn{1}{c|}{} & \multicolumn{1}{c|}{} & \multicolumn{1}{c|}{} & \multicolumn{1}{c|}{} & \multicolumn{1}{c|}{} & \multicolumn{1}{c|}{} & \multicolumn{1}{c|}{} & \multicolumn{1}{c|}{} & \multicolumn{1}{c|}{} & \multicolumn{1}{c|}{} & \multicolumn{1}{c|}{} & \multicolumn{1}{c|}{} & \multicolumn{1}{c|}{} \\
& \multicolumn{1}{c|}{} & \multicolumn{1}{c|}{} & \multicolumn{1}{c|}{} & \multicolumn{1}{c|}{} & \multicolumn{1}{c|}{} & \multicolumn{1}{c|}{} & \multicolumn{1}{c|}{} & \multicolumn{1}{c|}{} & \multicolumn{1}{c|}{} & \multicolumn{1}{c|}{} & \multicolumn{1}{c|}{} & \multicolumn{1}{c|}{} & \multicolumn{1}{c|}{} \\
& \multicolumn{1}{c|}{} & \multicolumn{1}{c|}{} & \multicolumn{1}{c|}{} & \multicolumn{1}{c|}{} & \multicolumn{1}{c|}{} & \multicolumn{1}{c|}{} & \multicolumn{1}{c|}{} & \multicolumn{1}{c|}{} & \multicolumn{1}{c|}{} & \multicolumn{1}{c|}{} & \multicolumn{1}{c|}{} & \multicolumn{1}{c|}{} & \multicolumn{1}{c|}{} \\
& \multicolumn{1}{c|}{} & \multicolumn{1}{c|}{} & \multicolumn{1}{c|}{} & \multicolumn{1}{c|}{} & \multicolumn{1}{c|}{} & \multicolumn{1}{c|}{} & \multicolumn{1}{c|}{} & \multicolumn{1}{c|}{} & \multicolumn{1}{c|}{} & \multicolumn{1}{c|}{} & \multicolumn{1}{c|}{} & \multicolumn{1}{c|}{} & \multicolumn{1}{c|}{}  \\
& \multicolumn{1}{c|}{} & \multicolumn{1}{c|}{} & \multicolumn{1}{c|}{} & \multicolumn{1}{c|}{} & \multicolumn{1}{c|}{} & \multicolumn{1}{c|}{} & \multicolumn{1}{c|}{} & \multicolumn{1}{c|}{} & \multicolumn{1}{c|}{} & \multicolumn{1}{c|}{} & \multicolumn{1}{c|}{} & \multicolumn{1}{c|}{} & \multicolumn{1}{c|}{}  \\
\hline
	
\multirow{6}{*}{\rotatebox{90}{Data Center}}
& \cite{eens_zong_2018} 
& E, H 
& Sim 
& \Circle 
& \CIRCLE 
& \Circle 
& \Circle 
& \CIRCLE 
& \CIRCLE 
& \CIRCLE 
& DA, SM
& NV
& {Energy saving with traffic grooming in optical DC networks} \\ \cline{2-14}
& \cite{eens_jahani_2019} 
& H 
& Sim 
&\Circle 
& \CIRCLE 
&\Circle 
&\Circle 
& \CIRCLE 
& \CIRCLE 
& \Circle 
& SM, EH
& NV
& {Resource reachability with renewable energy} \\ \cline{2-14}
& \cite{eens_pham_2020} 
& E, H 
& Sim, Exp  
& \Circle 
& \CIRCLE 
& \Circle 
& \Circle 
& \CIRCLE 
& \CIRCLE 
& \Circle 
& DA, SM
& NV
& {Considering router power consumption and network congestion} \\ \cline{2-14}
& \cite{eens_zhang_2020} 
& E, H 
& 
& \Circle 
& \CIRCLE 
& \Circle 
& \CIRCLE 
& \CIRCLE 
& \CIRCLE 
& \Circle 
& SM
& NV
& {Power consumption of network nodes with VN migration} \\ \cline{2-14}
& \cite{eens_kar_2018} 
& H 
& Sim 
& \CIRCLE 
& \CIRCLE 
& \LEFTcircle 
& \Circle 
& \CIRCLE 
& \CIRCLE 
& \Circle 
& DA, SM
& SV
& {Selecting more active nodes for the flow path; SFC} \\
\hline
\multirow{5}{*}{\rotatebox{90}{Transport}}
& \cite{eens_hejja_2018} 
& H 
& Sim 
& \Circle 
& \CIRCLE 
& \Circle 
& \Circle 
& \CIRCLE 
& \CIRCLE 
& \Circle 
& DA, SM
& NV
& {Segmentation of approach for substrate networks} \\ \cline{2-14}
& \cite{eens_chemodanov_2019} 
& E, H 
& Sim, Exp 
& \CIRCLE 
& \CIRCLE 
& \Circle 
& \Circle 
& \CIRCLE 
& \CIRCLE 
& \CIRCLE 
& DA
& NV, SV
& {Neighborhoods method for reducing cost of path traversal, including energy} \\ \cline{2-14}
& \cite{eens_he_2020} 
& S 
& Sim 
& \Circle
& \CIRCLE 
& \Circle 
& \Circle 
& \CIRCLE 
& \CIRCLE 
& \Circle 
& SM
& NV
& {Clustering method using dynamic regions of interest} \\ \cline{2-14}
& \cite{eens_cao_2020} 
&  H 
&  Sim 
&  \Circle 
&  \CIRCLE 
&  \Circle 
&  \Circle 
&  \CIRCLE 
&  \CIRCLE 
& \Circle 
&  DA
&  NV
& {Energy prices in multi-domain networks; VNE} \\ \cline{2-14}
& \cite{eens_huin_2018} 
& E, H 
& Sim 
& \CIRCLE 
& \CIRCLE 
& \LEFTcircle 
& \Circle 
& \CIRCLE 
& \CIRCLE 
& \CIRCLE 
& DA, SM
& SV
& {Consideration on active links and netw. interfaces; SFC} \\ \cline{2-14}
& \cite{eens_farkiani_2019} 
& H 
& Sim 
& \Circle 
& \CIRCLE 
& 
& 
& \CIRCLE 
& \CIRCLE 
& \Circle 
& DA, SM
& SV
& {Capacity of traffic processing, instance sharing; SFC} \\ \cline{2-14}
& \cite{eens_bari_2019} 
& E, H 
& Sim 
& \CIRCLE 
& \CIRCLE 
& 
& \CIRCLE 
& \CIRCLE 
& \CIRCLE 
& \CIRCLE 
& DA, SM
& SV
& {Minimizing carbon footprint; VNF-FGE and reconfig.} \\ \cline{2-14}
& \cite{eens_sun_2020} 
& H 
& Sim 
& 
& \CIRCLE 
& \Circle 
& \Circle 
& \CIRCLE 
& \CIRCLE 
& \Circle 
& DA
& SV
& {Reducing server energy consumption; SFC} \\ \cline{2-14}
& \cite{eens_moosavi_2021} 
& H 
& Sim 
& \CIRCLE 
& \CIRCLE
& \CIRCLE 
& \CIRCLE 
& \CIRCLE 
& \CIRCLE 
& \CIRCLE 
& DA, SM, \newline HT
& SV
& {Different modes of hybrid SDN/NFV networks; SFC} \\ \cline{2-14}
& \cite{eens_varasteh_2021} 
& E, S 
& Sim 
& \CIRCLE 
& \CIRCLE 
& \LEFTcircle 
& \Circle 
& \CIRCLE 
& \CIRCLE 
& \CIRCLE 
& DA, SM
& SV
& {Considering delay for SFC} \\ \cline{2-14}
& \cite{eens_pham_2021} 
& S 
& Sim 
& \CIRCLE 
& \CIRCLE 
& \Circle 
& \Circle 
& \CIRCLE 
& \Circle 
& \Circle 
& DA, SM \newline ML
& SV
& {Strict SFC latency; SFC} \\ \cline{2-14}
& \cite{eens_lin_2022} 
& H 
& Sim 
& \Circle 
& \CIRCLE 
& \CIRCLE 
& \Circle 
& \CIRCLE 
& \CIRCLE 
& \Circle 
& SM, HT
& SV
& {Nodes and optical-electronic networks; SFC} \\ \hline
\multirow{5}{*}{\rotatebox{90}{Wireless}}
& \cite{eens_ho_2019} 
& S 
& Sim 
& \Circle 
& \CIRCLE 
& \CIRCLE 
& \Circle 
& \CIRCLE 
& \CIRCLE 
& \Circle 
& DA, HT
& NV
& {Considering communications power of base stations} \\ \cline{2-14}
& \cite{eens_chen_2021} 
& S 
& Sim 
& \CIRCLE 
& \CIRCLE 
& \CIRCLE 
& \LEFTcircle 
& \CIRCLE 
& \CIRCLE 
& \CIRCLE 
& DA, HT
& NV
& {WLAN users' power consumption for communications} \\ \cline{2-14}
& \cite{eens_xu_2021} 
& S 
& Sim 
& \Circle 
& \CIRCLE 
& \Circle 
& \Circle 
& \CIRCLE 
& \CIRCLE 
& \Circle 
& DA, EH, \newline ML
& SV
& {Communications power and energy harvesting; Constrained RL} \\ \cline{2-14}
& \cite{eens_masoudi_2022} 
& S 
& Sim 
& \Circle 
& \CIRCLE 
& \Circle 
&\Circle 
& \CIRCLE 
& \CIRCLE 
& \Circle 
& DA, SM
& NV, SV
& {Networking, central cloud, midhaul/fronthaul, base stations, and network slices; E2E} \\ \cline{2-14}
& \cite{eens_taskou_2022} 
& H 
& Sim 
& \CIRCLE 
& \CIRCLE 
& \Circle 
& \Circle 
& \CIRCLE 
& \Circle 
& \Circle 
& DA
& SV
& {E2E with core, transport, backhaul, and RAN; SFC} \\
\hline
\multirow{5}{*}{\rotatebox{90}{Emerging}}
& \cite{eens_maity_2022} 
& E, H 
& Sim 
& \Circle 
& \CIRCLE 
& \Circle 
& \Circle 
& \CIRCLE 
& \CIRCLE 
& \Circle 
& SM
& NV
& {Limited maximum power for executing VNFs; VNE; Satellite} \\ \cline{2-14}
& \cite{eens_borylo_2020} 
& E, H 
& Sim 
& \CIRCLE 
& \CIRCLE 
& \CIRCLE 
& \Circle 
& \CIRCLE 
& \CIRCLE 
& \CIRCLE 
& DA, SM, \newline HT
& SV
& {Energy cost models for central and edge clouds; Cloud-edge} \\ \cline{2-14}
& \cite{eens_dawaliby_2019} 
& H 
& Sim 
& \CIRCLE 
& \CIRCLE 
& \Circle 
& \Circle 
& \CIRCLE 
& \CIRCLE 
& \CIRCLE 
& SM
& SV
& {Energy consumption of slices on a LoRa gateway; IoT} \\ \cline{2-14}
& \cite{eens_cappello_2022} 
& E, H 
& Sim 
& \CIRCLE 
& \CIRCLE 
& \LEFTcircle 
& \CIRCLE
& \CIRCLE 
& \CIRCLE 
& \CIRCLE 
& DA
& SV
& {Engines, containers, traffic manage., transmission; UAV} \\ \cline{2-14}
& \cite{eens_cao_2022} 
& H 
& Sim 
& \Circle 
& \CIRCLE 
& \Circle 
& \Circle 
& \CIRCLE 
& \CIRCLE 
& \Circle 
& SM
& SV
& {Energy costs of RAN, transport, and core; Vehicular} \\ \hline
\end{tabular}
\begin{tablenotes}
	\item E: Exact, H: Heuristic, S: Scheme, Sim: Simulation, Exp: Experimentation; 	DA: Dynamic Adaptation, SM: Sleep Modes, HT: Heterogeneous Resources, EH: Energy Harvesting; NV: Network Virtualization, SV: Service Virtualization
\end{tablenotes}
\end{table*}

\subsubsection{Data center networks}
Zong et al. \cite{eens_zong_2018} in this network scenario was on energy-aware VNE in optical networks, related to flexible-grid elastic; assuming sliceable transponders equipped in each node. Zhang et al. \cite{eens_zhang_2020} discussed virtual network migration in VNE due to changes in the physical network that affected the energy cost. Thus, the authors proposed migration techniques in the forms of virtual network algorithms to re-optimize energy cost. Zong et al. \cite{eens_zong_2018} attempted to reduce active DC and network components with location-aware VNE. Kar et al. \cite{eens_kar_2018} proposed a dynamic energy-saving model with NFV using M/M/c queueing network and minimum policy for starting a machine to reduce the frequent state changes.

\subsubsection{Transport networks}
Chemodanov et al. \cite{eens_chemodanov_2019} proposed exact and scheme approaches to optimize VNE with constrained shortest paths, also NFV, SFC, and traffic engineeering algorithms that led to network utilization and energy efficiency. The concurrent embedment problem was examined by Hejja et al. \cite{eens_hejja_2018}, considering SFC on split paths to utilize fragmented resources and minimize the number of open nodes. In \cite{eens_pham_2021}, the authors discuss a multi-objective VNE problem with congestion- and energy-aware, aimed at saving costs, energy, and avoiding network congestion simultaneously.

\subsubsection{Wireless networks}
This network scenario covers several NS related research paper, spanning from RAN to WSN. Xu et al. \cite{eens_xu_2021} proposed RAN slicing between MEC and traditional services. In \cite{eens_chen_2021}, the researchers provided a scheme and heuristics with slice resource allocation architecture for TV multimedia service in 5G \acs{C-RAN}. They used a DL approach for devising the solution. Joint optimization of mode selection and resource allocation in uplink fog RAN was discussed. Mobile virtual network operators (MNVO) topic were found in this scenario. The authors in \cite{eens_ho_2019} proposed a design framework for resource alocation in an orthogonal frequency-division multiple access (OFDMA) virtualized wireless network that met service contracts with differents MNVOs. Another framework was proposed by researchers Other topic, such as heterogeneity in wireless networks with \acs{D2D} communications was carried out by the authors. A resource allocation for Industry 4.0 was also investigated based on SDN, NFV, and \acs{ML} tools by Messaoud et al. \cite{eens_masoudi_2022}. Taskou et al. \cite{eens_taskou_2022} proposed a scheme for guaranteeing latency and reliability of sporadic ultra reliable low latency communications (URLLC) uplink traffic, while improving the quality of continuous enhanced mobile broadband (eMBB) services.

\subsubsection{Emerging networks}
Our survey in emerging networks scenario covers IoT, satellite, edge cloud, and vehicular networks. Maity et al. \cite{eens_maity_2022} discussed on improving diversity and flexibility of the service operation and management for IoT-powered CPS. Dawaliby et al. \cite{eens_dawaliby_2019} proposed a distributed slicing strategy for LoraWAN-based access networks with coalitional game and matching theory. A technique to model a business process as a virtual network considering virtual nodes representing the requested processing and locations was proposed by Cappello et al. \cite{eens_cappello_2022}.

\section{Future Research Challenges}
\label{section:frd}
In this section, future research challenges are discussed. We organized the points starting from programmable hardware to experimental evaluation environments.

\subsection{Green Programmable Hardware and Green Software}
To improve and realize "green programmable hardware," component manufacturers and hardware \acs{OEM}s are important. They have the role to design, produce, and supply components and equipments that comply to energy and softwarized network standards. This also can be forced by network/telecom operators or service providers \cite{mckinsey_greener_2020}, demanding improved energy-efficient network equipments/functions, and also include programmability supports when procuring. Hardware are commonly used with software, thus software OEMs are urged to do similar way, for example by following standards, tooling, and best practices from Green Software Foundation \cite{green_sf}.

\subsection{Energy Consumption in Control and MANO Layers} 
Network softwarization as the enabler for energy-efficient network infrastructure will provide control and MANO layers to accomplish the objective. However, the introduction of softwarized or virtualized layers that will provide global network view of resources, efficient use of resources, and isolated network services is also the source of additional energy consumption. Thus, researchers and industries need to consider the energy consumption in the control and MANO layers since it can be spanned from a server providing SDN controller to multiple machines for scalability, security, or reliability reasons. The energy models and strategies can be derived from general to specific computing and networking technologies.

\subsection{Network Reconfiguration and Sharing}
Network reconfiguration is a key management problem solved by softwarized networks. The feature will provide a way for network operators automatically reconfigure network infrastructure in a network scenario, including network topology, resources, functions, and traffic. The reconfiguration can be done to match demands or follow a certain objective, e.g., energy efficiency. Network scalability with scaling out and in for the components of the scenario (horizontal and vertical scaling). It also cover migration of virtual resources, functions, and traffic by re-routing to satisfy the actual target. Further, with the objective of efficient use of resources, including energy, network sharing is a technique that can be used from topology to VNFs.

\subsection{Energy Savings at the Edge}
Edge network is considered the place where all the future network innovation takes place in the coming years. Latency and energy are a few metrics that could be associated with users and end systems. Some of the use cases that will benefit from the edge are 5G massive machine-type communications (mMTC) and critical mMTC. In wired networks, optical fibers technology are being increasingly used to support flexibility by configuring and reconfiguring optical network devices, such as EON \cite{eesdn_hadi_2019}. However, common technology, like passive optical networks (PONs), can also be used to provide energy efficiency in the edge or access networks, and is still an improvement over electrical-based mediums such as copper.

There are three resources that need to be handled for obtaining energy savings at the edge: computing, networking, and storage/caching. Recently, researchers have been building solutions using network softwarization in cloud-edge/fog computing environments \cite{eesdn_luo_2020,eenfv_chaudhry_2020, eenfv_tinini_2019}, to jointly coordinate and orchestrate resources. It is important to realize that there could be billions of devices that potentially connected in 2030, with projection of 24 billion in 2025 \cite{gsma_intelligence_mobile_2021}, that will consume resources at the edge. Research on computing, networking including RAN, and storage/caching need to be jointly considered to achieve higher energy efficiency.

\subsection{E2E Energy-Efficient Softwarized Networks}
Researchers have been researching in energy efficiency in softwarized networks primarily in domains or parts of the network. There is still lack of investigation that considers E2E energy efficiency that includes end systems, wired/wireless network devices, and cloud/edge DCs. As 5G and beyond (5GB) networks are the network scenario that converge wired and wireless networks to be integrated with softwarization, use cases, and requirements, E2E aspect cannot be neglected. Thus, research with E2E energy efficiency using SDN/NFV/NS would be beneficial to obtain energy efficiency from one end, the consumer side, to the other end, the provider side.

The role of controller or MANO in SDN/NFV/NS is important to realize the E2E energy efficiency. However, since networks are interconnected to one another, the biggest one being the Internet, this is an auspicious goal. There are many operators and enterprises that own the networks, and they do not always agree to cooperate with one another. Hence, federation of network infrastructure, along with the trends of private 5GB networks, could be the solution to these issues targeting energy efficiency objective in E2E manner.

\subsection{Network Heterogeneity}
Network resources in a domain or multi-domain are commonly heterogeneous. This characteristic will be more and more pervasive in the emerging network scenarios in particular at the edge networks. It is not only related to physical network resources, but also the virtual ones. Starting from physical networks with different access networks, i.e., radio access networks (RAN) with multiple sizes of base stations, different wireless technologies, wired networks with different specifications, speeds, interfaces, intermediate nodes, and so on. With the emergence of programmable data plane, we also have challenges in the network accelerators targeting speeding up VNFs, packet processing, and also AI/ML computation in the data plane or infrastructure layer. Various types of network switches, SmartNICs, FPGAs, DPU/IPU, etc. with different specifications would also demand more attention.

\subsection{Energy Heterogeneity}
\begin{figure}
	\centering
	\includegraphics[scale=0.7]{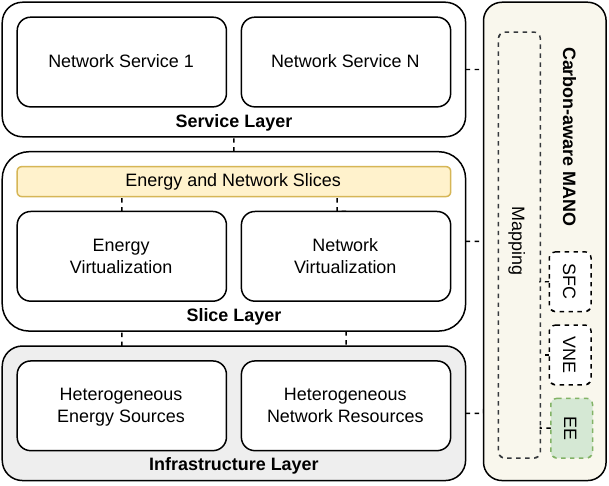}
	\caption{Sustainable services run on virtualized energy and network resources (slices), managed and orchestrated by a carbon-aware MANO.}
	\label{fig:sust_services}
\end{figure}

In energy sources, there are also similar challenges. Renewable and ambient energy sources with different kinds of energy generation, starting with wind and solar energy and other forms of renewable energy sources. Energy harvesting as one of the energy-efficient strategies can be used to get the ambient energy sources around network infrastructure, wireless networks in particular. This would also pose some challenges in energy management to the network infrastructure connected to these energy sources. With different types of energy sources from traditional electric grid, renewable, and ambient energy with energy storage or battery, we can virtualize these heterogeneous energy sources into energy virtualization \cite{Bashir2021}. Again, MANO layer that manages life cycle and orchestrate network resources would also be a help for integrating energy virtualization with compute, network, and storage virtualization to support and optimize sustainable services as described in {\figurename~{\ref{fig:sust_services}}}.

\subsection{Optimization Approaches}
Each network operator has their own objectives or criteria for designing and operating their networks. There are two approaches, traditional \cite{eesdn_maaloul_2018, eesdn_xie_2018, eenfv_kaur_2019, eesdn_hadi_2019, eesdn_zhao_2021, eesdn_kim_2020} and machine learning \cite{eesdn_chen_2020, eesdn_luo_2020,eenfv_xu_2020}. Sustainability as a common goal for many sectors and industries will drive prioritized objectives, including energy efficiency. However, seen in the literature review, this objective needs to be balanced with other criteria: QoS, scalability, heterogeneity, mobility, and federation; each pull against energy consumption. Hence, in optimizing this objective, researchers and industries need to consider multi-objective or pareto-optimal optimization \cite{Wang_2020}.

\subsection{Metrics and Measurements}
A handful of researchers have proposed metrics on energy efficiency in softwarized networks, e.g., RESDN \cite{eesdn_assefa_2020} and ECPUB \cite{eesdn_fang_2022}. These metrics were applied to different network scenarios and focusing on the energy per useful or utilized bits (capacity metric), instead of transmitted bit. These would lead to more metrics with different network scenarios. Thus, the traditional metric that uses bits/Joule needs to be updated or adapted based on the cases of the research or scenarios. Related to the current 5G or upcoming 6G networks, the KPIs would also need to consider this types of adaptation. However, the basics of the "new" metrics do not divert from the foundational standards. Related to the trends of zero carbon and carbon-aware techniques in various domains, energy metrics also need to be adapted to the carbon, since energy efficiency does not necessarily mean carbon reduction. Regarding to the heterogeneous energy sources---ambient and renewable energy sources in particular---, the integration of the measurements and calculation of the measured values have to be done correctly. Thus, the experimental evaluation methods as the realization of the research stages would be advantageous.

\subsection{Evaluation Environments}
Optimization techniques with simulations are common evaluation methods for energy savings with network softwarization research. Researchers often explored the problems from network design perspective, and simulate the problems using assumptions that could differ in their implementation in real-world situations. Several studies saw researchers focusing on problems with simulated scenarios, particularly with emulation techniques \cite{eens_huin_2018, eesdn_chen_2020}.

Energy-saving involving physical resources, such as power with voltage and current (components, circuit, electronic systems), is defined as electrical and needs real measurements. More experimentation for the evaluation using testbeds \cite{marzouk_energy_2020,eenfv_montazerolghaem_2020,eenfv_wu_2019, eens_chemodanov_2019} and real networks would be useful, in order to both realize solutions and encourage more research. Shared testbeds and experimentation environments with federation are highly needed to reduce costs in conducting research in this topic.\\

\section{Conclusions}
\label{section:conclusions}
Programmability and flexibility of network softwarization enable modern and future networks adaptive to dynamic changes. These features help softwarized networks in various and emerging scenarios to achieve operators' objectives, including energy efficiency, with the abstraction/virtualization provided by control and MANO layers. With softwarized network control by SDN, virtualized network functions by NFV, and isolated network resources by NS, network infrastructure can be managed by pushing energy-efficiency policies and orchestrating resources also services to support the objective via the MANO layer. This cannot be realized without energy efficiency optimization at the layer considering energy contributors, energy/power models, and energy efficiency strategies tailored (or can be adapted) to the actual network scenario.

This paper provided a survey on energy-efficient softwarized networks with supporting basics on energy efficiency in network infrastructure and three key technologies of network softwarization. Our major contributions are the classification of the literature considering network scenarios, approaches, criteria, metrics, and evaluation methods, as well as mapping the energy efficiency strategies and sublayers of each network softwarization technologies (control and MANO sublayers) to the classified papers. At last, we discussed potential research challenges awaited to be tackled regarding energy-efficient softwarized networks.

\bibliographystyle{IEEEtran}
\bibliography{IEEEabrv,ee_netsoft.bib}

\vfill
\vfill

\end{document}